\documentclass[aps,twocolumn,showpacs,preprintnumbers,footinbib,prl,10pt,superscriptaddress]{revtex4-1}

\makeatletter
\def\l@subsubsection#1#2{}
\def\l@subsubsubsection#1#2{}
\makeatother

\usepackage{graphicx,amssymb,amsmath,amsthm,amsfonts,epsfig,epsf,fixmath}
\usepackage[usenames]{color}
\usepackage{epstopdf}

\usepackage{bm}
\usepackage{dcolumn}
\usepackage{latexsym}
\usepackage{rotating}
\usepackage{longtable}

\usepackage{enumerate}
\usepackage{enumitem}
\usepackage{mathrsfs}
\usepackage{tensor,multirow}
\usepackage{url}
\usepackage{braket}
\usepackage{soul}
\usepackage[dvipsnames]{xcolor}
\usepackage[unicode]{hyperref}
\hypersetup{colorlinks=true,
            citecolor=MidnightBlue,
            linkcolor=MidnightBlue, urlcolor=MidnightBlue, linktocpage=true}
\usepackage{orcidlink}

\usepackage[normalem]{ulem}
\newcommand{\review}[2]{\ifmmode\text{\sout{\ensuremath{#1}}}\else\sout{#1}\fi\textcolor{BrickRed}{#2}}
\newcommand{\correct}[2]{\ifmmode\text{\sout{\ensuremath{#1}}}\else\sout{#1}\fi\textcolor{Fuchsia}{#2}}

\begin{document}

\title{Bayesian Search of Massive Scalar Fields from LIGO-Virgo-KAGRA Binaries}

\author{Yiqi~Xie~\orcidlink{0000-0002-8172-577X}}
\affiliation{Illinois Center for Advanced Studies of the Universe, Department of Physics, University of Illinois at Urbana-Champaign, Urbana, Illinois 61801, USA}

\author{Adrian~Ka-Wai~Chung~\orcidlink{0000-0003-2020-3254}}
\affiliation{Illinois Center for Advanced Studies of the Universe, Department of Physics, University of Illinois at Urbana-Champaign, Urbana, Illinois 61801, USA}

\author{Thomas~P.~Sotiriou~\orcidlink{0000-0002-9089-4866}}
\affiliation{Nottingham Centre of Gravity, University of Nottingham, University Park, Nottingham, NG7 2RD, United Kingdom}
\affiliation{School of Mathematical Sciences \& School of Physics and Astronomy, University of Nottingham, University Park, Nottingham, NG7 2RD, United Kingdom}

\author{Nicol\'as~Yunes~\orcidlink{0000-0001-6147-1736}}
\affiliation{Illinois Center for Advanced Studies of the Universe, Department of Physics, University of Illinois at Urbana-Champaign, Urbana, Illinois 61801, USA}

\begin{abstract} 

Massive scalar fields are promising candidates for addressing many unresolved problems in fundamental physics. We report the first model-agnostic Bayesian search of massive scalar fields that are nonminimally coupled to gravity in LIGO/Virgo/KAGRA gravitational-wave data. We find no evidence for such fields and place the most stringent upper limits on their coupling for scalar masses $\lesssim 2\times10^{-12}\,{\rm eV}$. We exemplify the strength of these bounds by applying them to massive scalar-Gauss-Bonnet gravity, finding the tightest constraints on the coupling constant to date, $\sqrt{\alpha_{\rm GB}}\lesssim 1\,{\rm km}$ for scalar masses $\lesssim 10^{-13}\,{\rm eV}$ to 90\% credible level.

\end{abstract}

\maketitle

\newcounter{footnote_suppmat}

\noindent{{\bf{\em Introduction.}}}
Scalar fields are ubiquitous in extensions of general relativity (GR) or the Standard Model of particle physics~\cite{PhysRevLett.40.223,Damour:1994zq,Arvanitaki:2009fg,Barack:2018yly,Berti:2015itd,Yunes:2024lzm}, motivated by the quest for quantum gravity and attempts to address internal consistency problems, {\em e.g.}~hierarchy problem~\cite{Arkani-Hamed:1998jmv}, strong {\em CP} problem~\cite{Hook:2018dlk}. Light scalars have also been suggested as potential explanations for dark energy~\cite{Copeland:2006wr} or dark matter~\cite{Hui:2016ltb,Hui:2021tkt}.

The inspiral behavior of compact binaries can be significantly affected if a scalar endows compact stars or black holes (BHs) with a scalar monopole, making gravitational waves (GWs) a promising probe of new fundamental scalars. Such a binary would emit scalar dipolar radiation, in addition to the standard quadrupole gravitational radiation. The extra loss of energy would affect the orbital dynamics and, in turn, the conventional gravitational-wave polarizations, leading to a GW dephasing that is ${\cal{O}}(v^2/c^2)$ larger than the leading-order term in the GR GW phase, where $v$ is the orbital velocity and $c$ is the speed of light [i.e.~a $-1$ post-Newtonian (PN) order effect]~\cite{Barausse:2016eii, Chamberlain:2017fjl, Alexander:2018qzg}. Thus, the effect of the additional dipolar emission on the orbital dynamics is very significant in the early inspiral phase. Indeed, pulsar observations have all but ruled out the prospect that compact stars could carry a scalar charge if the scalar field is massless ~\cite{Antoniadis:2013pzd,Anderson:2019eay}. LIGO-Virgo-KAGRA (LVK) observations have been used to search for scalar fields with black hole binaries~\cite{Yagi:2012gp,Nair:2019iur,Perkins:2021mhb,Gao:2024rel,Sanger:2024axs,Julie:2024fwy,Wang:2021jfc,Lyu:2022gdr,Wang:2023wgv,Saffer:2021gak}, while future observations of highly asymmetric binaries by the Laser Interferometer Space Antenna (LISA) have been shown to have great potential as well~\cite{Maselli:2020zgv, Maselli:2021men,Speri:2024qak}.

Searches for massive scalars with GW present an additional challenge: massive fields are confined near the compact objects and this suppresses scalar emission in the early inspiral. Indeed, GW observations are likely ``blind'' to fields whose inverse mass is smaller than the size of the compact objects of the binary. Nonetheless, axion-like particles are expected to be very light but not massless~\cite{PhysRevLett.40.223,Arvanitaki:2009fg,Hui:2021tkt}, while deviations from GR that lead to interesting non-linear strong field phenomena, such as scalarization~\cite{Damour:1993hw,Silva:2017uqg,Doneva:2017bvd,Dima:2020yac,Herdeiro:2020wei,Doneva:2022ewd}, include massive scalars. 
Indeed, the GW signal emitted by a binary neutron star (binary NS, or BNS), the GW170817 event, was analyzed for a model-specific search of axion at a discrete grid of axion mass in~\cite{Zhang:2021mks}. A model-agnostic search for massive scalar fields was explored in~\cite{Yamada:2019zrb} through Fisher analysis using a simulated, synthetic dataset.
Both analyses were done for some particular values of the scalar field mass only.

In this work, we report the first model-agnostic Bayesian search of massive scalar fields from GW signals detected by the LVK detectors during their first three observing runs (O1--O3)~\cite{LIGOScientific:2016dsl,LIGOScientific:2018mvr,LIGOScientific:2020ibl,LIGOScientific:2021usb,KAGRA:2021vkt,KAGRA:2023pio} as well as the latest one released in O4~\cite{LIGOScientific:2024elc}, without fixing the scalar mass $\mu_s\hbar$ \textit{a priori}. 
This analysis is agnostic to the theory that endows the binary component with a scalar charge. 
We find no evidence for dipolar emission in BH binaries (BNSs) 
for scalar masses $\mu_s\hbar\lesssim 5\times10^{-13}\,{\rm eV}$ ($2\times10^{-12}\,{\rm eV}$). In a more focused range $\mu_s\hbar\lesssim 3\times10^{-13}\,{\rm eV}$, we show constraints on the difference in the scalar charge per unit mass in BH binaries (BNSs) to be $\lesssim 0.5$ ($0.05$).
These are the first and most stringent Bayesian GW constraints on these massive scalar fields in this mass range. We note that dipole radiation is a generic feature and a smoking-gun effect of nonminimally coupled scalars on compact binaries, see e.g.~\cite{Alsing:2011er} for an example given in massive Brans-Dicke theory.
 
We also repeat our analysis for a specific massive scalar-tensor theory, massive scalar-Gauss-Bonnet (massive sGB, or msGB) gravity.
In geometric units $G=1=c$, the msGB action writes
\begin{align}
    S =&\; \int d^4x \sqrt{-g}\, \bigg[\frac{R}{16\pi} + \alpha_{\rm GB} \varphi \mathcal{X}_{\rm GB}^2 \notag\\ 
    &\;-\frac{1}{2}(\nabla_a\varphi\nabla^a\varphi + \mu_s^2\varphi^2) \bigg] + S_{\rm matter}, \label{eqn:sgb_action}
\end{align}
where $g$ is the determinant of the metric $g_{ab}$, $R$ is the Ricci scalar, $\varphi$ is a real scalar field with mass $\mu_s\hbar$,
$S_{\rm matter}$ is the matter action, $\mathcal{X}_{\rm GB}^2 = R^2 - 4 R_{ab} R^{ab} + R_{abcd} R^{abcd}$ is the Gauss-Bonnet (GB) invariant, $R_{ab}$ and $R_{abcd}$ are the Ricci tensor and the Riemann tensor, respectively and
$\alpha_{\rm GB}$ is a dimensionful coupling constant. 
We here obtain the first Bayesian LVK constraints on $\sqrt{\alpha_{\rm GB}}$ for a massive scalar, with 90\% credible intervals $\lesssim1\,\rm km$ for scalar masses $\lesssim10^{-13}\,\rm eV$. 

Massive sGB provides a rather minimal model for scalar hair from a massive field in BH binaries, and yet it is sufficiently general for inspiral modeling.  The linear coupling between $\varphi$ and $\mathcal{X}_{\rm GB}^2$ included in our action is the only shift-symmetric term that can evade no-hair theorems~\cite{Sotiriou:2011dz,Hui:2012qt,Sotiriou:2015pka} and lead to scalar hair~\cite{Yunes:2011we,Sotiriou:2013qea,Sotiriou:2014pfa,Ayzenberg:2014aka}. 
More general couplings with $\mathcal{X}_{\rm GB}^2$, couplings with other curvature invariants, or self-coupling of the scalar, could exhibit broader phenomenology in general ({\em e.g.}~\cite{Silva:2017uqg,Doneva:2017bvd,Dima:2020yac,Herdeiro:2020wei,Doneva:2022ewd}), but are expected to be subdominant in our setup and can be thus modeled perturbatively. 
A nonzero $\mu_s$ would be expected if such additional interactions break shift-symmetry, and it could also arise from non-perturbative instanton effects~\cite{Alexander:inprep}.
Hence, we consider massive sGB a fairly representative model for our analysis.

\vspace{0.5em}
\noindent{{\bf{\em Dipole emission from massive scalar fields.}}}
BHs with scalar hair in the $\mu_s=0$ case of the action in Eq.~\eqref{eqn:sgb_action} have been studied extensively~\cite{Yunes:2011we,Sotiriou:2013qea,Sotiriou:2014pfa,Prabhu:2018aun}.
The scalar asymptotes to $\varphi_{\rm BH}(r) = Q_{\rm BH}/r$, where $r$ is the radial coordinate and $Q_{\rm BH}$ is the scalar charge. At the leading order in $\alpha_{\rm GB}$~\cite{Berti:2018cxi,Saravani:2019xwx,Lyu:2022gdr},
\begin{align}
    Q_{\rm BH} = \frac{\sqrt{16\pi}\,\alpha_{\rm GB}}{m}\, \frac{2\sqrt{1-\chi^2}}{1+\sqrt{1-\chi^2}},
    \label{eqn:sgb_charge}
\end{align}
where $m$ and $\chi$ are the mass and the dimensionless spin of the BH, respectively. 
For $\mu_s>0$, there is an additional Yukawa-like suppression of the scalar field at large distances, and the asymptotic field profile takes the form $\varphi_{\rm BH}(r) = Q_{\rm BH}\,e^{-\mu_s r}/r$.
Here, we approximate $Q_{\rm BH}$ endowed by the massive field with Eq.~\eqref{eqn:sgb_charge}, which is valid because GW observations only allow for small $\mu_s$ values to be searched over (see~\cite{vanGemeren:2024bzf} for further discussion of scalar hair in msGB). 
Note that a linear coupling between a scalar and the GB invariant does not endow neutron stars (NSs) with a scalar monopole \cite{Yagi:2015oca}, and hence, hereafter we take $Q_{\rm NS}=0$.

When at least one of the two compact objects in a binary carries scalar charge there will be dipolar emission but it will only kick in when the orbital angular frequency, $\Omega$, reaches the Compton angular frequency of the scalar field.  
Indeed, for quasicircular orbits, the dipole radiation power from a generic massive scalar field has been solved in~\cite{Krause:1994ar,Alexander:2018qzg},
\begin{align}
    \delta\dot{E} = -\frac{1}{3} \eta^2 M^2 \Omega^4 r_{12}^2\,|\Delta\tilde{Q}|^2 \left(1-\frac{\mu_s^2}{\Omega^2}\right)^{3/2},~\label{eqn:dipole_power}
\end{align}
where subscripts $1,2$ denote the primary and secondary component of the binary, $\Delta\tilde{Q}=Q_1/m_1-Q_2/m_2$ is a dimensionless dipole parameter, $M=m_1+m_2$ is the total mass, $\eta=m_1m_2/M^2$ is the symmetric mass ratio, $r_{12}$ is the orbital separation.

The change of radiation power in Eq.~\eqref{eqn:dipole_power} can be mapped to the inspiral waveform using the parametrized post-Einsteinian framework~\cite{Yunes:2009ke,Yunes:2016jcc} (see the Supplemental Material~\footnote{See Supplemental Material, which includes Refs.~\cite{Hannam:2013oca,Husa:2015iqa,Khan:2015jqa,Pratten:2020ceb,Pratten:2020fqn,Garcia-Quiros:2020qpx,Dietrich:2019kaq,Chatziioannou:2012rf,Mezzasoma:2022pjb,Alexander:2018qzg,Yunes:2009ke,Yunes:2016jcc,lalsuite,Mehta:2022pcn,Lyu:2022gdr,LIGOScientific:2019lzm,KAGRA:2023pio,LIGOScientific:2018mvr,LIGOScientific:2020ibl,LIGOScientific:2021usb,KAGRA:2021vkt,LIGOScientific:2020stg,LIGOScientific:2020zkf,LIGOScientific:2021qlt,LIGOScientific:2024elc,LIGOScientific:2018mvr,LIGOScientific:2020ibl,LIGOScientific:2021usb,KAGRA:2021vkt,Planck:2015fie,DES:2017kbs,Cantiello:2018ffy,LIGOScientific:2018hze,Levan:2017ubn,Hjorth:2017yza}, for details about the modified gravitational waveform, GW events selected for the analysis, parameter estimation scheme and postprocessing procedure, individual posterior for each GW event, and analysis of posterior artifacts that leads to exclusion of certain GW events from the discussion in the main text.}\setcounter{footnote_suppmat}{\value{footnote}} for details). 
We focus here on the dominant $(2,2)$ harmonic, as others are related through a simple scaling~\cite{Chatziioannou:2012rf,Mezzasoma:2022pjb,Mehta:2022pcn}. In the frequency domain, the modified waveform can be written as
\begin{align}
    \tilde{h}(f) = \tilde{h}_{\rm GR}(f)\, e^{-i\delta\Psi(f)}, \label{eqn:ppe}
\end{align}
where $f=\Omega/\pi$ is the GW frequency, $\tilde{h}_{\rm GR}$ is a GR waveform, which we here choose from the \texttt{IMRPhenom} family ({\em e.g.}~\cite{Hannam:2013oca,Husa:2015iqa,Khan:2015jqa,Pratten:2020ceb,Pratten:2020fqn,Garcia-Quiros:2020qpx,Dietrich:2019kaq}), and $\delta\Psi$ describes the correction to GR. At a stage where the dipole has been activated but the binary is far from merging, the correction is
\begin{align}
    \delta\Psi(f) \sim -\frac{5\,|\Delta\tilde{Q}|^2}{7168\,\eta\,(\pi Mf)^{7/3}},\quad f_{\rm act} < f < f_{\rm insp}, \label{eqn:delta_psi}
\end{align}
up to a linear function of $f$, where $f_{\rm act}=\mu_s/\pi$ is the dipole activation frequency and $f_{\rm insp}=0.018/M$ is an estimated ending frequency of the inspiral.
Beyond the frequency range prescribed above, we apply no physical modification but only linearly extrapolate Eq.~\eqref{eqn:delta_psi} to satisfy the requirement of $C^1$ continuity. We neglect corrections from the change in the binary's binding energy as, for a massive field, these would affect higher PN orders only~\cite{Alexander:2018qzg}. 

\vspace{0.5em}
\noindent{{\bf{\em Gravitational wave parameter estimation.}}}
We use LVK open data~\cite{LIGOScientific:2019lzm,KAGRA:2023pio} and focus on specific events selected for the LVK parametrized inspiral tests of GR~\cite{LIGOScientific:2020tif,LIGOScientific:2021sio,LIGOScientific:2018dkp}, each of which is (i) detected by at least two detectors, (ii) has a false-alarm rate less than $10^{-3}\,{\rm yr}^{-1}$, and (iii) accumulates an SNR greater than $6$ during the inspiral. 
We further filter the list with the requirement that either $M<30\,{\rm M_\odot}$ or there is strong evidence of mass asymmetry by the LVK analysis~\cite{LIGOScientific:2020stg,LIGOScientific:2020zkf,LIGOScientific:2021qlt}, as a smaller $M$ elongates the inspiral, while, according to Eq.~\eqref{eqn:delta_psi}, the same dipole $|\Delta\tilde{Q}|$ leads to a greater GR deviation when $M$ and $\eta$ are smaller. 
We also add the latest O4 event, GW230529~\cite{LIGOScientific:2024elc}, for the msGB analysis since several works before us~\cite{Gao:2024rel,Sanger:2024axs,Julie:2024fwy} have claimed the tightest constraint on massless sGB gravity using this event. 
This leaves us with 16 events (see Supplemental Material~\footnotemark[\value{footnote_suppmat}] for a full list).

For each event, we perform Bayesian parameter estimation with the waveform model of Eqs.~\eqref{eqn:ppe}--\eqref{eqn:delta_psi} and a Gaussian noise model. 
The $\tilde{h}_{\rm GR}$ function in Eq.~\eqref{eqn:ppe} is taken to be \texttt{IMRPhenomPv2}~\cite{Hannam:2013oca,Husa:2015iqa,Khan:2015jqa} for symmetric BH binaries and \texttt{IMRPhenomPv2\_NRTidalv2}~\cite{Dietrich:2019kaq} for BNSs. For asymmetric binaries or NSBH candidates, we choose $\tilde{h}_{\rm GR}$ to be \texttt{IMRPhenomXPHM}~\cite{Pratten:2020ceb,Pratten:2020fqn,Garcia-Quiros:2020qpx} with an additional $(3,3)$ mode that reasonably covers higher-multipole contributions~\cite{LIGOScientific:2020zkf,LIGOScientific:2020stg,LIGOScientific:2021qlt}.

Exploiting the \texttt{Bilby} inference library~\cite{Ashton:2018jfp} with the \texttt{dynesty} nested sampler~\cite{Speagle:2019ivv}, we estimate the posterior distribution for $\vec{\lambda}_{\rm GR}\cup\{\mu_s,|\Delta\tilde{Q}|\textrm{ or }\sqrt{\alpha_{\rm GB}}\}$, 
where $\vec{\lambda}_{\rm GR}$ are source parameters, such as binary masses and spins, which $\tilde{h}_{\rm GR}$ depends on. 
The posterior is then marginalized over $\vec{\lambda}_{\rm GR}$ and smoothed with a method based on Gaussian kernel density estimation.
We extract the 90\% bounds for $|\Delta\tilde{Q}|$ and $\sqrt{\alpha_{\rm GB}}$ as functions of $\mu_s$ when constraining a generic massive dipole and msGB gravity, respectively. 
In the latter case, the waveform is reparametrized with Eq.~\eqref{eqn:sgb_charge}, and a combined 90\% bound is also obtained by multiplying the likelihoods from the single-event analysis.

Because NSs do not acquire scalar charges in sGB gravity, we only analyze BNSs when constraining a generic massive dipole. 
The source of GW190814 may either be a binary BH or an NSBH~\cite{LIGOScientific:2020zkf}. The former case would result in a tighter sGB constraint because then $|\Delta\tilde{Q}|$ would be dominated by $\alpha_{\rm GB}/m_s^2$, where $m_s$ is the smaller mass in the binary that carries a scalar charge; the dipole effect is smaller for NSBHs because $m_s$ has to take the larger, primary (BH) mass value.
Here, we assume GW190814 is an NSBH to obtain a conservative msGB constraint.
The source of GW230529 is an NS and an object of unknown nature (most likely a BH)~\cite{LIGOScientific:2024elc}. 
Following~\cite{Gao:2024rel,Sanger:2024axs,Julie:2024fwy}, we analyze this event as an NSBH, but we do not include it in the combined msGB analysis.

Our waveform model assumes that the modification to GR is small so we must check that this is respected by our posteriors.
In~\cite{Nair:2019iur,Perkins:2021mhb}, $\sqrt{\alpha_{\rm GB}}/m_s<0.5$ is proposed as a validity cutoff in \textit{massless} sGB theory.
We will not try to enforce this condition, or attempt to generalize it here, as the $\mu_s>0$ case is fundamentally different. Since dipolar emission effectively kicks in at a certain frequency, one can have virtually zero deviations from GR for parts of the waveform, even for large couplings, while deviations could still be significant in the later inspiral. Hence, we will instead require 
\begin{align}
\mathcal{N}_e(\delta\Psi)/\mathcal{N}_e(\Psi_{\rm GR}^{0{\rm PN}})<1, \label{eqn:validity_wf}
\end{align}
where $\Psi_{\rm GR}^{0{\rm PN}}$ is the leading-PN-order GR phase, while 
\begin{align}
\mathcal{N}_e(\Psi)=\min_{\phi,t}\left[\int\frac{df\,|\tilde{h}(f)|^2(\Psi(f)+\phi+2\pi ft)^2}{4\pi^2\,{\rm SNR}^2\,S_n(f)}\right]^{1/2}\,,
\end{align}
are \textit{effective cycles}~\cite{Sampson:2014qqa}, which measure the number of GW cycles incurred by $\Psi$, as weighted by a noise spectral density $S_n$. 
According to condition~\eqref{eqn:validity_wf}, the validity cutoff for $\sqrt{\alpha_{\rm GB}}$ (or $|\Delta\tilde{Q}|$) increases with $\mu_s$, because the frequency range of modifications in $\delta\Psi$ is smaller when $\mu_s$ is larger. 

Given Eq.~\eqref{eqn:delta_psi}, when $\mu_s>\pi f_{\rm insp}$, there is no modification to GR, and a constraint on the dipole is not possible. 
In the Supplemental Material~\footnotemark[\value{footnote_suppmat}], we show that the likelihood does not die out fast enough as  $|\Delta\tilde{Q}|\rightarrow\infty$, but rather it asymptotes to the GR likelihood multiplied by a factor equal to the GR posterior probability of $f_{\rm insp}<\mu_s/\pi$. 
We refer to the latter as the ``fraction of indifference'' (FOI) and only report the 90\% bounds in the range ${\rm FOI}(\mu_s)<1\%$, where the GR posterior for computing the FOI is estimated with a separate nested sampling run under the same computational settings. 
For the combined msGB constraint, we multiply the FOIs in alignment with the multiplication of single-event likelihoods.

The priors used in our analysis are uniform over $|\Delta\tilde{Q}|\in[0,1]$ and $\sqrt{\alpha_{\rm GB}}\in[0,10]\,{\rm km}$, except that for GW230529 the $\sqrt{\alpha_{\rm GB}}$ prior is limited within $[0,1]\,{\rm km}$ to avoid unphysical results falsely supported by the single-detector data~\cite{Sanger:2024axs}. 
The prior over $\mu_s$ is uniform in a logarithmic scale, between $f_{\rm act}=\mu_s/\pi=10\,{\rm Hz}$ and a $\mu_s$ value sufficiently beyond the FOI limit. 
The lower bound of $\mu_s$ can be treated in the same way as in the $\mu_s\rightarrow0$ limit because the LVK data has a lower frequency cutoff at $20\,{\rm Hz}$ and it is blind to any activation before that.
The prior choice for $\vec{\lambda}_{\rm GR}$ is adapted from the LVK standard analysis assuming GR~\cite{LIGOScientific:2018mvr,LIGOScientific:2020ibl,LIGOScientific:2021usb,KAGRA:2021vkt,Planck:2015fie}, and combines electromagnetic observations when available~\cite{Levan:2017ubn,Hjorth:2017yza,DES:2017kbs,Cantiello:2018ffy,LIGOScientific:2018hze}.
After running parameter estimation, we find 3 events showing posterior artifacts that are irrelevant to the massive scalar field of interest, and we exclude these events from results (both single-event and combined) presented hereafter.
See Supplemental Material~\footnotemark[\value{footnote_suppmat}] for a full list of $\vec{\lambda}_{\rm GR}$ and their priors, the detailed settings and postprocessing for each sampling run, the posterior results of each event, and discussion of the posterior artifacts.

\begin{figure*}[ht]
    \centering
    \includegraphics[width=0.98\textwidth]{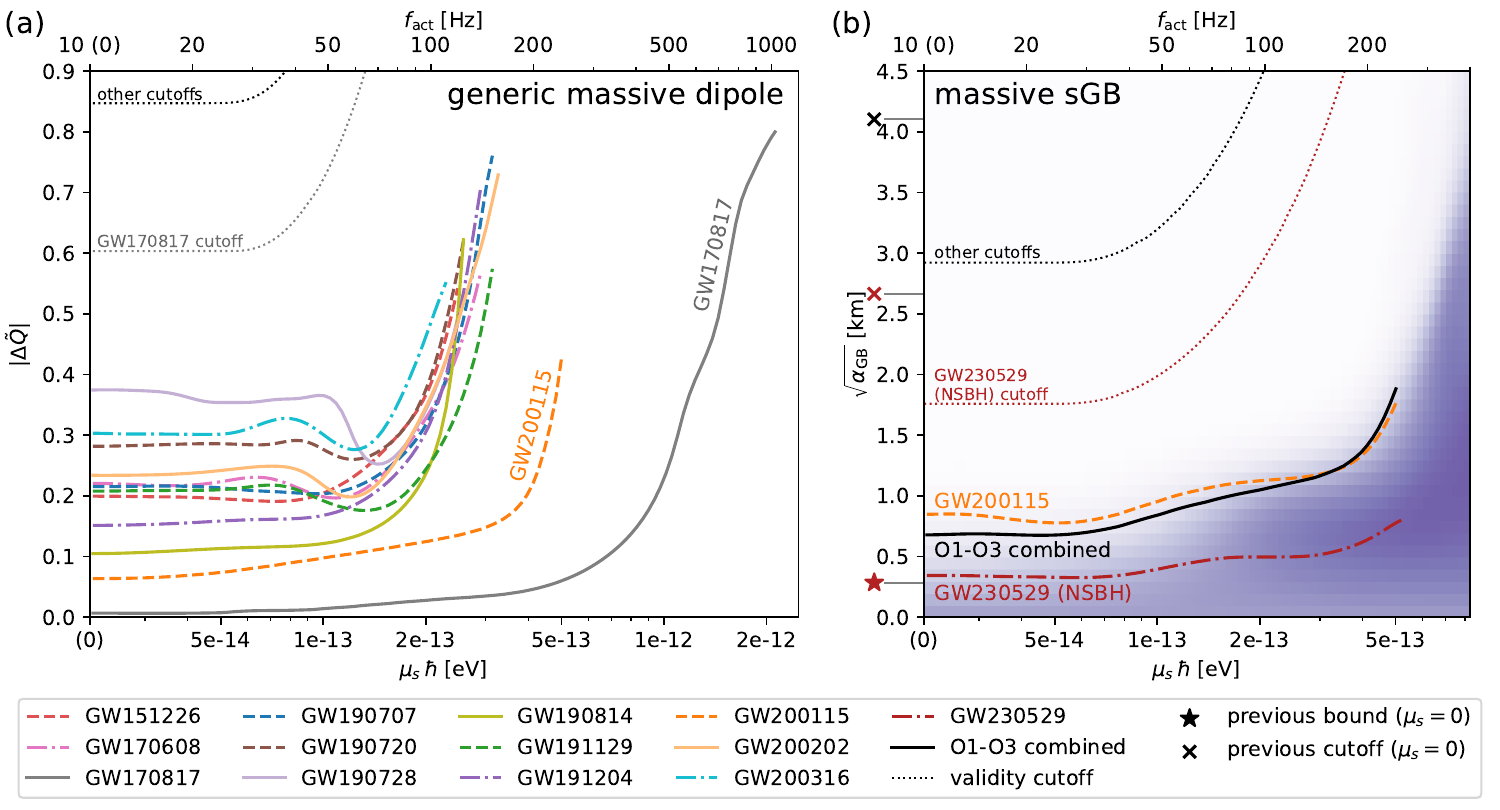}
    \caption{
    Constraints on dipole emission from massive scalar fields. 
    Panel (a) shows the 90\% bound on the dimensionless dipole parameter $|\Delta\tilde{Q}|$ as a function of the scalar mass $\mu_s$ for each O1--O3 binary. Panel (b) shows the same as (a), but for the sGB coupling constant $\sqrt{\alpha_{\rm GB}}$ in msGB gravity, using the GW200115 event and the combination of O1--O3 BH binaries (accompanied by the combined 2D posterior shaded in purple). Panel (b) is also overlaid with the bound from the O4 event GW230529 assuming it is an NSBH, which is not included in the combined analysis.
    The left end of each bound curve is equivalent to the $\mu_s\rightarrow0$ limit, given the lower-frequency cutoff of the data. 
    The right end is limited by the FOI condition or the finite width of the $|\Delta\tilde{Q}|$ or $\sqrt{\alpha_{\rm GB}}$ prior [GW170817 in panel (a) and GW230529 in panel (b)].
    The dotted curves show the validity cutoffs for specific binaries, while ``other cutoffs'' refer to the lowest one across O1--O3 BH binaries.
    Each star (cross) marks the center of previous 90\% bounds~\cite{Gao:2024rel,Sanger:2024axs,Julie:2024fwy} for massless sGB (consistent with a $\sqrt{\alpha_{\rm GB}}/m_s<0.5$ cutoff~\cite{Nair:2019iur,Perkins:2021mhb}), with different colors indicating different binaries, i.e.~crimson for GW230529 and black for ``others'' combined.
    }
    \label{fig:main_constraints}
\end{figure*}
\vspace{0.5em}
\noindent{{\bf{\em Constraints on dipole emission from massive scalars.}}}
In Fig.~\ref{fig:main_constraints} (a), we show the single-event constraints on $|\Delta\tilde{Q}|$ for generic dipole emission. 
A black dotted curve is additionally drawn for the lowest validity cutoff across all O1--O3 BH binaries, and we confirm that every 90\% bound from such a binary is below this curve (and hence also below each one's own validity cutoff). The same also applies to the GW170817 and GW230529 constraints, for which the validity cutoffs are significantly lower and are plotted separately.
As previously discussed, the constraint becomes weaker and is eventually entirely lifted as $f_{\rm act}=\mu_s/\pi$ approaches and surpasses $f_{\rm insp}$, where we stop reporting the 90\% bound based on the ${\rm FOI}$ condition. 
For GW170817 and GW230529, however, we observe a saturation of the prior before the ${\rm FOI}$ limit, so we only show the range in which the posterior 90\% bound is below 80\% of the prior maximum.
Despite that, GW170817 still presents the widest constraints, $\mu_s\hbar\lesssim2.1\times10^{-12}\,{\rm eV}$, because it has the smallest binary total mass, and hence, it is the longest inspiral---for the same reason we also observe the most stringent dipole constraint $|\Delta\tilde{Q}|\lesssim0.0067$ in the $\mu_s\rightarrow0$ limit from GW170817.

Between the $\mu_s\rightarrow0$ limit and the point where $|\Delta\tilde{Q}|$ becomes practically unconstrained, the variation of the 90\% bound is nonmonotonic for each event. 
Sometimes, the bound on $|\Delta\tilde{Q}|$ can at first become tighter as one moves to larger  $\mu_s$, before becoming looser again. 
This is because the sudden activation of the dipole would be better observed by the detectors as $f_{\rm act}$ moves toward their sensitivity buckets, near $\mathcal{O}(10^2)\,{\rm Hz}$. 

In Fig.~\ref{fig:main_constraints} (b), we show the combined constraint on $\sqrt{\alpha_{\rm GB}}$ in msGB gravity, together with the combined posterior (shaded in purple) and the best single-event constraint from GW200115 up to O3. 
These constraints satisfy condition \eqref{eqn:validity_wf} and $\sqrt{\alpha_{\rm GB}}/m_s<0.5$ by their lowest cutoffs across all events. 
Because FOIs are multiplied, the combined constraint is reported up to a higher scalar mass $\mu_s\hbar\lesssim5.2\times10^{-13}\,{\rm eV}$, where we see the weakest 90\% bound is $\sqrt{\alpha_{\rm GB}}\lesssim1.88\,{\rm km}$.
In the massless limit $\mu_s\rightarrow0$, the combined 90\% bound suggests $\sqrt{\alpha_{\rm GB}}\lesssim0.68\,{\rm km}$~\footnote{Previous works~\cite{Lyu:2022gdr,Wang:2023wgv} have reported a different 90\% bound of $\sqrt{\alpha_{\rm GB}}\lesssim1\,{\rm km}$ for massless sGB using O1--O3 binaries. However, we note a sign error in the non-GR dephasing term of the waveform model used by (at least)~\cite{Lyu:2022gdr}.}
Our results show that such a constraint (to 90\% credible level) is maintained all the way up to $\mu_s\hbar\lesssim10^{-13}\,{\rm eV}$, before significantly weakening.

We have not included the O4 event GW230529~\cite{LIGOScientific:2024elc} in our combined analysis as it could well be a BNS, in which case the dipole emission would be suppressed~\cite{Yagi:2015oca}. 
However, since Refs.~\cite{Gao:2024rel,Sanger:2024axs,Julie:2024fwy} have used this event to obtain tight constraints on massless sGB assuming the source is an NSBH, we have analyzed it as a single event under the same assumption and overlaid its single-event msGB 90\% bound curve on top of Fig.~\ref{fig:main_constraints} (b). In the $\mu_s\rightarrow0$ limit, we find $\sqrt{\alpha_{\rm GB}}\lesssim0.34\,{\rm km}$, which is consistent with~\cite{Gao:2024rel,Sanger:2024axs,Julie:2024fwy}. 
The full curve, therefore, shows how these results are extended to the massive regime (up to $\mu_s\hbar\lesssim5.2\times10^{-13}\,{\rm eV}$ due to the prior edge, see Supplemental Material~\footnotemark[\value{footnote_suppmat}])

\vspace{0.5em}
\noindent{{\bf{\em Discussion and future prospects.}}}
We have conducted the first model-agnostic Bayesian search of massive scalar fields nonminimally coupled to gravity using LVK data and also considered a specific well-motivated theory, msGB gravity. In both cases, we find no modifications to GR, and, when the mass is below a certain threshold (set by the characteristic frequency of the binary), we obtain constraints on the scalar charge or coupling that are as stringent as those for a massless scalar, but for a wide range of masses. The constraints on the charge or coupling constant are comparable for all scalar masses below the threshold, strongly suggesting that imposing a bound on or measuring a mass below that threshold will be hard.  Past that threshold, bounds on the charge or coupling weaken rapidly. 

The tightest constraints from a single event come from different events in each of the two approaches. The BNS event that gives the tightest bound in the model-agnostic search is entirely absent in the msGB case, as NSs do not carry scalar monopoles in this theory. But even for BH binaries, one gets the tightest bounds from different events. In msGB, where one can meaningfully combine events, the resulting constraint is a significant improvement with respect to the tightest single-event constraints, at least for lower masses. 

For lower masses, we expect the constraints we have obtained in msGB gravity to be conservative and robust to the inclusion of additional interaction terms, provided that the linear coupling between the scalar and the GB invariant is the dominant contribution to the scalar charge for stationary BHs. Additional interaction terms would then be expected to contribute at higher PN orders~\cite{Shiralilou:2021mfl}, which would be relevant only in the late inspiral. 

The above discussion is based on current LVK data, but future observations of lighter binaries and longer inspirals with more advanced ground-based detectors~\cite{KAGRA:2013rdx,Punturo:2010zz,Dwyer:2014fpa} can progressively improve the constraints. Pushing the bounds to higher scalar masses is challenging though. As the scalar mass gets closer to the threshold beyond which the constraints on the charge rapidly weaken, the onset of dipolar emission is pushed to the late inspiral, and higher PN corrections become more relevant, in both a model-agnostic and a theory-specific approach. In addition, the baseline GR model also becomes less accurate approaching the merger. Hence, obtaining reliable bounds in that part of the parameter space will be rather challenging both technically and observationally.

Space-based detectors planned in the 2030s~\cite{LISA:2017pwj,Sato:2017dkf,TianQin:2015yph,Hu:2017mde} will open up a new window for constraining scalar fields with extreme mass-ratio inspirals (EMRIs)~\cite{Maselli:2020zgv,Maselli:2021men,Barsanti:2022vvl}.
Current conservative estimates suggest that detecting EMRIs with LISA would yield a bound on the order of $Q_2/m_2\sim 0.002$ or $\sqrt{\alpha_{\rm GB}}\sim 0.2\,{\rm km}$~\cite{Speri:2024qak}. EMRI observations have also been shown to be able to measure or provide a bound for the scalar mass~\cite{Barsanti:2022vvl}. However, LISA bounds are inherently limited to lower scalar masses than LVK or 3G detectors, as they probe larger separations.
An interesting prospect, if nonzero scalar charges were to be detected, is combining observations from ground-based detectors and LISA to place a bound on the mass of the scalar.  

\begin{acknowledgments}
\vspace{0.5em}
\noindent{{\bf{\em Acknowledgments.}}}
We thank Carl-Johan~Haster, Anna~Liu, Hector~O.~Silva, and Kent~Yagi for useful discussions.
Y.X., A.K.W.C., and N.Y. acknowledge support from the Simons Foundation through Award No.~896696, the NSF through Grant No.~PHY-2207650 and NASA through Grant No.~80NSSC22K0806. 
Y.X. also acknowledges support from the Illinois Center for Advanced Studies of the Universe (ICASU) / Center for AstroPhysical Surveys (CAPS) Graduate Fellowship.
T.P.S. acknowledges partial support from the STFC Consolidated Grant No.~ST/V005596/1 and No.~ST/X000672/1.  
This work made use of the Illinois Campus Cluster, a computing resource that is operated by the Illinois Campus Cluster Program (ICCP) in conjunction with the National Center for Supercomputing Applications (NCSA), and is supported by funds from the University of Illinois Urbana-Champaign (UIUC).
\end{acknowledgments}

\bibliographystyle{utphys}
\bibliography{Ref}

\clearpage
\appendix
\section*{Supplemental Material}

\subsection{Gravitational waveform for dipole emission from a massive scalar field}

In the following, we derive the frequency-domain waveform of the $(\ell,{\rm m})$ harmonic with modifications to GR in the phase,
\begin{align}
    \tilde{h}_{\ell{\rm m}}(f) = \tilde{h}_{\rm GR}^{\ell{\rm m}}(f)\, e^{-i\delta\Psi_{\ell{\rm m}}(f)}. \label{eqn:ppe_all_harmonics}
\end{align}
The waveform presented in the main text corresponds to the $(\ell,{\rm m})=(2,2)$ mode, although we do add the $(\ell,{\rm m})=(3,3)$ mode when considering very asymmetric binaries.

The GW model for the coalescence of compact binaries can be piecewise decomposed in the frequency domain into an inspiral model, an intermediate model (related to the plunge), and a merger-ringdown model. Schematically, this can be written as
\begin{align}
    \tilde{h}_{\ell{\rm m}}(f) =&\; A_{\ell{\rm m}}(f)\, e^{-i\Psi_{\ell{\rm m}}(f)}, \notag\\
    A_{\ell{\rm m}}(f) =&\; \left\{\begin{array}{ll}
        A_{\ell{\rm m}}^{\rm insp}(f), &f<f_{\rm insp}, \\
        A_{\ell{\rm m}}^{\rm int}(f), &f_{\rm insp}\leq f<f_{\rm int}, \\
        A_{\ell{\rm m}}^{\rm mr}(f), &f\geq f_{\rm int},
    \end{array}\right. \notag\\
    \Psi_{\ell{\rm m}}(f) =&\; \left\{\begin{array}{ll}
        \Psi_{\ell{\rm m}}^{\rm insp}(f), &f<f_{\rm insp}, \\
        \Psi_{\ell{\rm m}}^{\rm int}(f), &f_{\rm insp}\leq f<f_{\rm int}, \\
        \Psi_{\ell{\rm m}}^{\rm mr}(f), &f\geq f_{\rm int},
    \end{array}\right. \label{eqn:imr_scheme}
\end{align}
where ``insp'', ``int'' and ``mr'' stand for inspiral, intermediate and merger-ringdown, respectively. The three pieces are separated by the inspiral ending frequency $f_{\rm insp}$ and an intermediate ending frequency $f_{\rm int}$, where the neighboring pieces are matched by requiring $C^1$ continuity\footnote{The actual values chosen for $f_{\rm insp}$ and $f_{\rm int}$ can be slightly different between the amplitude model and the phase model, and we note that Eq.~\eqref{eqn:imr_scheme} is just for schematically explaining the idea behind IMRPhenom models.}. 
This is the case for all \texttt{IMRPhenom} waveforms~\cite{Hannam:2013oca,Husa:2015iqa,Khan:2015jqa,Pratten:2020ceb,Pratten:2020fqn,Garcia-Quiros:2020qpx,Dietrich:2019kaq} which we choose for $\tilde{h}_{\rm GR}^{\ell{\rm m}}$, and hence we also model the modification $\delta\Psi_{\ell{\rm m}}$ following the same behavior. 

Let us focus on the inspiral model first. As pointed out in~\cite{Chatziioannou:2012rf,Mezzasoma:2022pjb}, modifications to different harmonics in the inspiral are related through the scaling,
\begin{align}
    \delta\Psi^{\rm ins}_{\ell{\rm m}}(f) \sim {\rm m}\,\Phi(2\pi f/{\rm m}), \label{eqn:harmonic_scaling}
\end{align}
up to a linear function of $f$ depending on some time and phase of reference, and $\Phi$ is a function we will define below. The correction to the GW inspiral phase in the frequency domain, for example, due to dipole emission, is sourced by the dipole correction to the energy fluxes (for a quasi-circular orbit), 
\begin{align}
    \dot{E}_{\rm GR} =&\; -\frac{32}{5} \eta^2 M^2 \Omega^6 r_{12}^4, \\
    \delta \dot{E} =&\; -\frac{1}{3} \eta^2 M^2 \Omega^4 r_{12}^2\,|\Delta\tilde{Q}|^2 \left(1-\frac{\mu_s^2}{\Omega^2}\right)^{3/2},
\end{align}
where $\dot{E}_{\rm GR}$ is the energy flux in GR and $\delta \dot{E}$ is the dipole correction during the inspiral. 
At leading-PN order, $r_{12}=(M\Omega)^{1/3}$ by Kepler's third law, and the $(1-\mu_s^2/\Omega^2)^{3/2}$ behavior can be approximated with a Heaviside step function $\Theta(\Omega - \mu_s)$~\cite{Alexander:2018qzg}.
Following the ppE formalism~\cite{Yunes:2009ke,Yunes:2016jcc}, the energy fluxes enter the phase modification in the inspiral through an integral in the stationary phase approximation\footnote{The sign here depends on the sign convention of Eq.~\eqref{eqn:ppe_all_harmonics} and of the phase in $\tilde{h}_{\rm GR}^{\ell{\rm m}}$. In this paper, $\tilde{h}_{\rm GR}^{22}\sim \exp[-i(3/128\eta)(\pi Mf)^{-5/3}]$ at the leading PN order, which follows from the \texttt{LALSimulation} implementation~\cite{lalsuite} of the \texttt{IMRPhenom} waveforms (the implementation adopted by \texttt{Bilby}).},
\begin{align}
    &\Phi(\Omega) = -\frac{5}{96\,\eta\,M^{5/3}} \int^\Omega d\Omega'\, \frac{\Omega-\Omega'}{\Omega^{\prime 11/3}} \frac{\delta\dot{E}}{\dot{E}_{\rm GR}} \notag\\
    &= -\frac{5\,|\Delta\tilde{Q}|^2}{14336\,\eta}\, \bigg[\frac{\Theta(\Omega - \mu_s)}{(M \Omega)^{7/3}} + \frac{\Theta(\mu_s - \Omega)}{3\,(M \mu_s)^{7/3}}\left(10-\frac{7\Omega}{\mu_s}\right)\bigg]. \label{eqn:ppe_aux_func}
\end{align}

In the intermediate and merger-ringdown stage, we do not add new modifications to GR, as our model for the dipole correction is only valid for the inspiral. However, due to the requirement of $C^1$ continuity, $\delta\Psi^{\rm ins}_{\ell{\rm m}}$ still impacts the later stages of coalescence through its ending value and its derivative. Also, recalling that Eq.~\eqref{eqn:harmonic_scaling} is left with an unspecified linear function of $f$, the entire inspiral-merger-ringdown correction model takes the form
\begin{align}
    \delta\Psi_{\ell{\rm m}}(f)
    =&\; {\rm m}\,\Phi(2\pi f/{\rm m})\,\Theta(f_{\rm insp} - f) \notag\\
    &\; + {\rm m}\,\Phi(2\pi f_{\rm insp}/{\rm m})\, \Theta(f - f_{\rm insp}) \notag\\
    &\; + 2\pi (f-f_{\rm insp})\,\Phi'(2\pi f_{\rm insp}/{\rm m})\, \Theta(f - f_{\rm insp}) \notag\\
    &\; + \Phi^{(0)}_{\ell{\rm m}} + \Phi^{(1)}_{\ell{\rm m}} f. \label{eqn:waveform_final}
\end{align}
The numbers $\Phi^{(0)}_{\ell{\rm m}}$ and $\Phi^{(1)}_{\ell{\rm m}}$ are determined such that the reference phase and the time of arrival of $\tilde{h}_{\rm GR}$ are not modified, i.e.
\begin{align}
    \delta\Psi_{\ell{\rm m}}({\rm m}f_{\rm ref}/2)=&\;0, \label{eqn:waveform_phi_fix} \\
    \delta\Psi'_{\ell{\rm m}}(f_{\rm peak})=&\;0, \label{eqn:waveform_t_fix}
\end{align}
where $f_{\rm ref}$ and $f_{\rm peak}$ are the reference frequency and peak frequency of $\tilde{h}_{\rm GR}^{22}$, respectively (cf.~\cite{Mehta:2022pcn}). 
We note that $f_{\rm peak}$ typically occurs after the inspiral, and given the linear continuation in Eq.~\eqref{eqn:waveform_final}, we replace $f_{\rm peak}$ with $f_{\rm insp}$ in Eq.~\eqref{eqn:waveform_t_fix} to simplify the actual implementation with
\begin{align}
    \delta\Psi'_{\ell{\rm m}}(f_{\rm insp})=0.\label{eqn:waveform_t_fix_equiv}
\end{align}
Finally, we choose
\begin{align}
    f_{\rm insp}=0.018/M, \label{eqn:f_insp}
\end{align}
following~\cite{Lyu:2022gdr}, which is also the inspiral ending frequency for the phase of \texttt{IMRPhenomPv2}~\cite{Hannam:2013oca,Husa:2015iqa,Khan:2015jqa}. 

To summarize, our waveform model is built by modifying the GR model with Eq.~\eqref{eqn:ppe_all_harmonics}. The master equation for the modification is Eq.~\eqref{eqn:waveform_final}, where the $\Phi$ function is given in Eq.~\eqref{eqn:ppe_aux_func}, the numbers $\Phi^{(0)}_{\ell{\rm m}}$ and $\Phi^{(1)}_{\ell{\rm m}}$ are determined by solving Eqs.~\eqref{eqn:waveform_phi_fix} and \eqref{eqn:waveform_t_fix_equiv}, and the inspiral ending frequency $f_{\rm insp}$ takes the value in Eq.~\eqref{eqn:f_insp}.
We note that the physical modification to GR in our waveform is characterized by
\begin{align}
    \delta\Psi_{\ell{\rm m}}(f) \sim -\frac{5\,{\rm m}\,|\Delta\tilde{Q}|^2}{14336\,\eta\,(\pi Mf)^{7/3}},\; f_{\rm act}^{\ell{\rm m}} < f < f_{\rm insp}, \label{eqn:delta_psi_lm}
\end{align}
where $f_{\rm act}^{\ell{\rm m}}=({\rm m}/2\pi)\mu_s$ is the dipole activation frequency in the $(\ell,{\rm m})$ mode.

\subsection{Formulation of parameter estimation}
The formulation of GW parameter estimation follows from Bayes' theorem,
\begin{align}
    p(\vec{\lambda}|\tilde{s}) = \frac{\mathcal{L}(\tilde{s}|\vec{\lambda})\, p(\vec{\lambda})}{\mathcal{Z}(\tilde{s})},\quad
    \mathcal{Z}(\tilde{s}) = \int \mathcal{L}(\tilde{s}|\vec{\lambda})\, p(\vec{\lambda})\, d\vec{\lambda},
\end{align}
where $p(\vec{\lambda}|\tilde{s})$ is the posterior distribution of parameters $\vec{\lambda}$ given the frequency-domain strain data $\tilde{s}$, $\mathcal{L}(\tilde{s}|\vec{\lambda})$ is the likelihood of obtaining the data from the model with a given set of parameters, $p(\vec{\lambda})$ is the prior distribution of the parameters, and $\mathcal{Z}(\tilde{s})$ is the evidence for the model.

Let us first consider single-event analysis. For the $i$th event with data $\tilde{s}^{(i)}$, the model parameters can be decomposed into
\begin{align}
    \vec{\lambda}^{(i)} = \vec{\lambda}_{\rm GR}^{(i)}\cup\{\mu_s,\epsilon^{(i)}\},
\end{align}
where $\vec{\lambda}_{\rm GR}^{(i)}$ are parameters required by the GR waveform model $\tilde{h}_{\rm GR}$, and $\epsilon^{(i)}=|\Delta\tilde{Q}^{(i)}|$ for constraining a generic dipole or $\sqrt{\alpha_{\rm GB}}$ for constraining massive sGB gravity. 
Let us not specify the prior at the moment, but we note that $\vec{\lambda}_{\rm GR}^{(i)}$ and $\{\mu_s,\epsilon^{(i)}\}$ should be independent of each other, so
\begin{align}
    p(\vec{\lambda}^{(i)})=p(\vec{\lambda}_{\rm GR}^{(i)})\,p(\mu_s,\epsilon^{(i)}).
\end{align}
Assuming that the noise is additive and Gaussian, the likelihood function is constructed as
\begin{align}
    \mathcal{L}(\tilde{s}^{(i)}|\vec{\lambda}^{(i)}) \propto e^{-\frac{1}{2} \langle \tilde{s}^{(i)}-\tilde{h}(\vec{\lambda}^{(i)})|\tilde{s}^{(i)}-\tilde{h}(\vec{\lambda}^{(i)})\rangle_{n^{(i)}}},
\end{align}
with the noise-weighted inner product
\begin{align}
    \langle\tilde{A}|\tilde{B}\rangle_n \equiv 4{\rm Re}\int_{f_{\rm low}}^{f_{\rm high}} \frac{\tilde{A}^*(f)\,\tilde{B}(f)}{S_{n}(f)}\,df,
\end{align}
where $S_n$ is the one-sided detector power spectral density, and $[f_{\rm low},f_{\rm high}]$ mark the frequency range of the data.

Once the posterior is thoroughly explored [for example through nested sampling or Markov-Chain Monte Carlo (MCMC) methods] and Bayesian parameter estimation is done, we extract the marginalized posterior for $\{\mu_s,\epsilon^{(i)}\}$ and its conditional form via
\begin{align}
    p(\mu_s,\epsilon^{(i)}|\tilde{s}^{(i)}) =&\; \int p(\vec{\lambda}^{(i)}|\tilde{s}^{(i)})\, d\vec{\lambda}_{\rm GR}^{(i)}, \\
    p(\epsilon^{(i)}|\mu_s,\tilde{s}^{(i)}) =&\; \frac{p(\mu_s,\epsilon^{(i)}|\tilde{s}^{(i)})}{\int p(\mu_s,\epsilon^{(i)}|\tilde{s}^{(i)})\, d\epsilon^{(i)}}. \label{eqn:posterior_conditional}
\end{align}
The 90\% bound curve for $\epsilon^{(i)}$ is then solved for from the conditional probability
\begin{align}
    P(\epsilon^{(i)}|\mu_s,\tilde{s}^{(i)}) = \int_0^{\epsilon^{(i)}}\!\! p(\epsilon^{(i)\prime}|\mu_s,\tilde{s}^{(i)})\, d\epsilon^{(i)\prime} = 0.9. \label{eqn:posterior_bound}
\end{align}
Another quantity that will be useful in later discussion is the marginalized likelihood, 
\begin{align}
    \mathcal{L}(\tilde{s}^{(i)}|\mu_s,\epsilon^{(i)}) = \int \mathcal{L}(\tilde{s}^{(i)}|\vec{\lambda}^{(i)})\, p(\vec{\lambda}_{\rm GR}^{(i)})\, d\vec{\lambda}_{\rm GR}^{(i)}, \label{eqn:likelihood_marginalized}
\end{align}
which can be thought of as an intermediate step towards the marginalized posterior,
\begin{align}
    p(\mu_s,\epsilon^{(i)}|\tilde{s}^{(i)}) \propto \mathcal{L}(\tilde{s}^{(i)}|\mu_s,\epsilon^{(i)})\, p(\mu_s,\epsilon^{(i)}).
\end{align}

When constraining massive sGB gravity, we will also stack data across events $\{\tilde{s}\}$. Formally, this is done by constructing the combined likelihood as the product of all single-event likelihoods,
\begin{align}
    \mathcal{L}(\{\tilde{s}\}|\{\vec{\lambda}_{\rm GR}\},\mu_s,\sqrt{\alpha_{\rm GB}})
    = \prod_i \mathcal{L}(\tilde{s}^{(i)}|\vec{\lambda}_{\rm GR}^{(i)},\mu_s,\sqrt{\alpha_{\rm GB}}).
\end{align}
and the combined posterior follows from
\begin{align}
    &p(\{\vec{\lambda}_{\rm GR}\},\mu_s,\sqrt{\alpha_{\rm GB}}|\{\tilde{s}\}) \notag\\
    &\propto p(\{\vec{\lambda}_{\rm GR}\},\mu_s,\sqrt{\alpha_{\rm GB}})\, \mathcal{L}(\{\tilde{s}\}|\{\vec{\lambda}_{\rm GR}\},\mu_s,\sqrt{\alpha_{\rm GB}}),
\end{align}
Because each source is believed to be independent from every other source, the prior can be decomposed as
\begin{align}
    p(\{\vec{\lambda}_{\rm GR}\},\mu_s,\sqrt{\alpha_{\rm GB}}) = p(\mu_s,\sqrt{\alpha_{\rm GB}}) \prod_i p(\vec{\lambda}_{\rm GR}^{(i)}).
\end{align}

In practice, we are only interested in the combined marginalized posterior. Using the prior decomposition,
\begin{align}
    &p(\mu_s,\sqrt{\alpha_{\rm GB}}|\{\tilde{s}\}) \notag\\
    &= \int p(\{\vec{\lambda}_{\rm GR}\},\mu,\sqrt{\alpha_{\rm GB}}|\{\tilde{s}\}) \prod_i d\vec{\lambda}_{\rm GR}^{(i)} \notag\\
    &\propto p(\mu_s,\sqrt{\alpha_{\rm GB}})^{-(N_{\rm event}-1)}\, \prod_i p(\mu_s,\sqrt{\alpha_{\rm GB}}|\tilde{s}^{(i)}),
\end{align}
where $N_{\rm event}$ is the total number of events. 
In coordinates where $p(\mu_s,\sqrt{\alpha_{\rm GB}})$ is transformed to a uniform distribution, the combined posterior is just the product of all single-event posteriors. 
One may also write
\begin{align}
    p(\mu_s,\sqrt{\alpha_{\rm GB}}|\{\tilde{s}\}) \propto p(\mu_s,\sqrt{\alpha_{\rm GB}})\, \mathcal{L}(\{\tilde{s}\}|\mu_s,\sqrt{\alpha_{\rm GB}}),
\end{align}
where the combined marginalized likelihood is given by
\begin{align}
    &\mathcal{L}(\{\tilde{s}\}|\mu_s,\sqrt{\alpha_{\rm GB}}) \notag\\
    &= \int \mathcal{L}(\{\tilde{s}\}|\{\vec{\lambda}_{\rm GR}\},\mu_s,\sqrt{\alpha_{\rm GB}}) \prod_i p(\vec{\lambda}_{\rm GR}^{(i)})\, d\vec{\lambda}_{\rm GR}^{(i)} \notag\\
    &= \prod_i \mathcal{L}(\tilde{s}^{(i)}|\mu_s,\sqrt{\alpha_{\rm GB}}).
\end{align}
The combined 90\% bound can be extracted by applying the same operations in Eqs.~\eqref{eqn:posterior_conditional} and \eqref{eqn:posterior_bound} to $p(\mu_s,\sqrt{\alpha_{\rm GB}}|\{\tilde{s}\})$.

\subsection{Quality of constraints and fraction of indifference}
According to Eq.~\eqref{eqn:delta_psi_lm}, the dipole modification vanishes when $f_{\rm act}^{\ell{\rm m}}>f_{\rm insp}$ for all harmonic modes, or equivalently when
\begin{align}
    \mu_s>(2\pi/{\rm m}_{\rm low})f_{\rm insp}, \label{eqn:region_indiff}
\end{align}
where ${\rm m}_{\rm low}$ is the lowest harmonic number ${\rm m}$ involved. This means that, at sufficiently large $\mu_s$, the dipole cannot be constrained. 

Under the Bayesian framework, the quality of the constraint can be assessed using the marginalized likelihood. 
In each single-event analysis, in order for the 90\% bound to be valid, we expect $\mathcal{L}(\tilde{s}^{(i)}|\mu_s,\epsilon^{(i)})$ to die out sufficiently fast as $\epsilon^{(i)}$ approaches its prior boundary -- if this is not the case, then we should increase the prior range.
This motivates us to investigate
\begin{align}
    &\lim_{\epsilon^{(i)}\rightarrow\infty} \mathcal{L}(\tilde{s}^{(i)}|\mu_s,\epsilon^{(i)}) \notag\\
    &=\lim_{\epsilon^{(i)}\rightarrow\infty} \left(\int_{\rm I} + \int_{\rm II}\right) \mathcal{L}(\tilde{s}^{(i)}|\vec{\lambda}^{(i)})\, p(\vec{\lambda}_{\rm GR}^{(i)})\, d\vec{\lambda}_{\rm GR}^{(i)},
\end{align}
where we split the $\vec{\lambda}_{\rm GR}^{(i)}$ space into two regions, ${\rm I}$ and ${\rm II}$, based on whether the condition of Eq.~\eqref{eqn:region_indiff} is satisfied.
In region ${\rm II}$, the modification is on, and supposedly the integrand completely dies out as $\epsilon^{(i)}\rightarrow\infty$. In region ${\rm I}$, however, the integrand is always equivalent to the GR likelihood ($\epsilon^{(i)}=0$). We can then write
\begin{align}
    &\lim_{\epsilon^{(i)}\rightarrow\infty} \mathcal{L}(\tilde{s}^{(i)}|\mu_s,\epsilon^{(i)}) \notag\\
    &=\int_{\rm I} \mathcal{L}(\tilde{s}^{(i)}|\vec{\lambda}_{\rm GR}^{(i)},\mu_s,\epsilon^{(i)}=0)\, p(\vec{\lambda}_{\rm GR}^{(i)})\, d\vec{\lambda}_{\rm GR}^{(i)} \notag\\
    &=\mathcal{L}(\tilde{s}^{(i)}|\mu_s,\epsilon^{(i)}=0) \int_{\rm I} p(\vec{\lambda}_{\rm GR}^{(i)}|\mu_s,\epsilon^{(i)}=0,\tilde{s}^{(i)})\, d\vec{\lambda}_{\rm GR}^{(i)} \notag\\
    &=\mathcal{L}(\tilde{s}^{(i)}|\mu_s,\epsilon^{(i)}=0)\, P({\rm I}|\mu_s,\epsilon^{(i)}=0,\tilde{s}^{(i)}).\label{eqn:likelihood_die_out}
\end{align}
Note that between the second and the third lines above, we have again applied Bayes' theorem,
\begin{align}
    p(\vec{\lambda}_{\rm GR}^{(i)}|\mu_s,\epsilon^{(i)}=0,\tilde{s}^{(i)}) =&\; \frac{\mathcal{L}(\tilde{s}^{(i)}|\vec{\lambda}_{\rm GR}^{(i)},\mu_s,\epsilon^{(i)}=0)\, p(\vec{\lambda}_{\rm GR}^{(i)})}{\mathcal{L}(\tilde{s}^{(i)}|\mu_s,\epsilon^{(i)}=0)}, \notag\\
    \mathcal{L}(\tilde{s}^{(i)}|\mu_s,\epsilon^{(i)}=0) =&\; \int \mathcal{L}(\tilde{s}^{(i)}|\vec{\lambda}_{\rm GR}^{(i)},\mu_s,\epsilon^{(i)}=0) \notag\\
    &\; \quad \times p(\vec{\lambda}_{\rm GR}^{(i)})\, d\vec{\lambda}_{\rm GR}^{(i)}.
\end{align}
Equation~\eqref{eqn:likelihood_die_out} means that, no matter how much we increase the prior range, $\mathcal{L}(\tilde{s}^{(i)}|\mu_s,\epsilon^{(i)})$ is smaller than $\mathcal{L}(\tilde{s}^{(i)}|\mu_s,\epsilon^{(i)}=0)$ by a factor of $P({\rm I}|\mu_s,\epsilon^{(i)}=0,\tilde{s}^{(i)})$ at most, and constraints on $\epsilon^{(i)}$ are reasonable only at those $\mu_s$ values where this factor is small enough (e.g.~$<1\%$).

In the main text, the factor $P({\rm I}|\mu_s,\epsilon^{(i)}=0,\tilde{s}^{(i)})$ is referred to as the fraction of indifference (FOI), because region ${\rm I}$ corresponds to a place where the waveform is indifferent to the presence of the scalar field. 
As suggested by Eq.~\eqref{eqn:likelihood_die_out}, the FOI can be obtained from a parameter estimation run assuming GR and the same $\vec{\lambda}_{\rm GR}$ prior, i.e.
\begin{align}
    {\rm FOI}(\mu_s|\tilde{s}^{(i)}) = P({\rm I}|\mu_s,\epsilon^{(i)}=0,\tilde{s}^{(i)}) = P_{\rm GR}({\rm I}|\tilde{s}^{(i)}),
\end{align}
where $P_{\rm GR}({\rm I}|\tilde{s}^{(i)})$ is the GR posterior probability of getting the condition in Eq.~\eqref{eqn:region_indiff} satisfied by the source parameters. 

When multiple events are stacked for constraining massive sGB gravity, Eq.~\eqref{eqn:likelihood_die_out} becomes
\begin{align}
    &\lim_{\sqrt{\alpha_{\rm GB}}\rightarrow\infty} \mathcal{L}(\{\tilde{s}\}|\mu_s,\sqrt{\alpha_{\rm GB}}) \notag\\
    &=\mathcal{L}(\{\tilde{s}\}|\mu_s,\sqrt{\alpha_{\rm GB}}=0)\, \prod_i P_{\rm GR}({\rm I}|\tilde{s}^{(i)}),
\end{align}
because the combined likelihood is the product of all single-event likelihoods. This means that the FOI is also multiplied, with
\begin{align}
    {\rm FOI}(\mu_s|\{\tilde{s}\}) = \prod_i P_{\rm GR}({\rm I}|\tilde{s}^{(i)}).
\end{align}

\subsection{Computational settings for parameter estimation}

The events analyzed in this work are explicitly listed in Table~\ref{tab:events}.
We load strain data from the Gravitational Wave Open Science Center~\cite{LIGOScientific:2019lzm,KAGRA:2023pio}, and follow the same choice of signal duration, frequency range, noise spectral density estimates and glitch mitigation described in~\cite{LIGOScientific:2018mvr,LIGOScientific:2020ibl,LIGOScientific:2021usb,KAGRA:2021vkt}. Given these settings, we note that $f_{\rm low}\geq20\,{\rm Hz}$ for all strains. 

\begingroup
\setlength{\tabcolsep}{5pt}
\begin{table}[htbp]
    \centering
    \begin{tabular}{lccc}
        \hline
        Event identifier & Binary type & $M$ [$M_\odot$] & Asym. \\
        \hline
        GW151226 & BBH & 21.7 & -- \\
        GW170608 & BBH & 18.5 & -- \\
        GW170817 & BNS & 2.7 & -- \\
        GW190412 & BBH & 36.8 & \checkmark~\cite{LIGOScientific:2020stg} \\
        GW190707\_093326 & BBH & 20.1 & -- \\
        GW190720\_000836 & BBH & 21.8 & -- \\
        GW190728\_064510 & BBH & 20.7 & -- \\
        GW190814 & BBH/NSBH & 25.9 & \checkmark~\cite{LIGOScientific:2020zkf} \\
        GW190924\_021846 & BBH & 13.9 & -- \\
        GW191129\_134029 & BBH & 17.5 & -- \\
        GW191204\_171526 & BBH & 20.2 & -- \\
        GW191216\_213338 & BBH & 19.8 & -- \\
        GW200115\_042309 & NSBH & 7.4 & \checkmark~\cite{LIGOScientific:2021qlt} \\
        GW200202\_154313 & BBH & 17.6 & -- \\
        GW200316\_215756 & BBH & 21.2 & -- \\
        GW230529\_181500 & NSBH(?) & 5.1 & -- \\
        \hline
    \end{tabular}
    \caption{Events selected for our analysis. The columns are, in order, the event identifier, the most likely type of source binary, the total mass by the LVK median estimate, and whether the components are evidently asymmetric. The identifiers will be shortened in later reference by dropping the information after the underscores. A question mark is left for the GW230529 source because the nature of the primary mass is unknown, though NSBH appears to be the most probable solution~\cite{LIGOScientific:2024elc}.
    }
    \label{tab:events}
\end{table}
\endgroup

For BH binaries before the GW230529 event, we choose \texttt{IMRPhenomPv2} and \texttt{IMRPhenomXPHM} as the base GR waveform $\tilde{h}_{\rm GR}$ for symmetric and asymmetric masses, respectively. In the latter case, an additional $(3,3)$ mode is added on top of the dominant $(2,2)$ mode. These GR waveforms are parametrized by
\begin{align}
    \vec{\lambda}_{\rm GR} = \{m_1, m_2, \vec{\chi}_1, \vec{\chi}_2, t_c, \phi_{\rm ref}, \psi, \iota, \alpha, \delta, D_L\},
\end{align}
where $m_{1,2}$ are the component masses, $\vec{\chi}_{1,2}$ and the component dimensionless spin vectors, $t_c$ is the coalescence time, $\phi_{\rm ref}$ is the reference phase at the reference frequency, $\psi$ is the polarization angle, $\iota$ is the inclination angle, $\alpha$ is the right ascension, $\delta$ is the declination, and $D_L$ is the luminosity distance. 

Similar to the LVK analysis~\cite{LIGOScientific:2018mvr,LIGOScientific:2020ibl,LIGOScientific:2021usb,KAGRA:2021vkt}, we choose a uniform prior over the redshifted component masses, spin magnitudes, coalescence time and reference phase, and an isotropic prior over the spin orientation, binary orientation and sky location. 
In particular, the prior over the masses is limited by $m_2/m_1\in[0.125,1]$ for \texttt{IMRPhenomPv2} and $[0.05,1]$ for \texttt{IMRPhenomXPHM}. 
The prior over the spin magnitudes ranges in $[0,0.99]$, for both BHs and NSs.
The prior over the coalescence time is restricted to $\pm0.1\,{\rm s}$ around the trigger time of the event.
For the luminosity distance, we choose a prior that is uniform in the source frame volume. A $\Lambda$-CDM cosmology with $H_0=67.9\,{\rm km}\,{\rm s}^{-1}{\rm Mpc}^{-1}$ and $\Omega_{\rm m}=0.3065$~\cite{Planck:2015fie} is assumed to compute the redshift, as well as the prior over the luminosity distance. 

In the beyond-GR sector, we choose uniform priors over $|\Delta\tilde{Q}|\in[0,1]$ and $\sqrt{\alpha_{\rm GB}}\in[0,10]\,{\rm km}$ (except that $\sqrt{\alpha_{\rm GB}}\in[0,1]\,{\rm km}$ for GW230529). For $\mu_s$, we choose a prior uniform in the logarithmic scale, bounded by $f_{\rm act}^{22}\in[10,500]\,{\rm Hz}$. The massless limit, $\mu_s\rightarrow0$, is covered as long as the redshifted $f_{\rm act}^{\ell{\rm m}}$ is smaller than $f_{\rm low}$ by the left bound of the prior, which is also true when the $(3,3)$ mode is involved.

In order to estimate the posteriors, we perform nested sampling using \texttt{Bilby} with the \texttt{dynesty} sampler. 
Each parameter estimation runs with 1000 live points and stops at \texttt{dlogz=0.1}. The MCMC evolution in each nested sampling step is done with the \texttt{Bilby}-implemented \texttt{acceptance-walk} method, with evolution length controlled by \texttt{naccept=60} when the GR base waveform is \texttt{IMRPhenomPv2} or \texttt{naccept=100} when the GR base waveform is \texttt{IMRPhenomXPHM}.
We further repeat each parameter estimation 10 times with different random seeds and combine the samples to improve the resolution. 

For the GW230529 event, the settings are mostly similar to those asymmetric BH binaries above. However, in order to account for the longer inspiral of GW230529, we sample over a wider $\mu_s$ range in $f_{\rm act}^{22}\in[10,800]\,{\rm Hz}$. 
Instead of repeating the same parameter estimation run with different seeds, we improve the sampling resolution by combining multiple runs with different segments of the full $\mu_s$ prior range. In particular, we carry out four runs with $f_{\rm act}^{22}\in[10,100]\,{\rm Hz}$, $[100,200]\,{\rm Hz}$, $[200,400]\,{\rm Hz}$, and $[400,800]\,{\rm Hz}$, respectively. 
When combining different samples, the sample weights are assigned through 
$w^{[i]}N^{[i]}\propto \mathcal{Z}^{[i]}\mathcal{V}^{[i]}$, where superscript $[i]$ is the index of the run, $w$ is the weight to be assigned, $N$ is the sample size, $\mathcal{Z}$ is the evidence which is also given by nested sampling, and $\mathcal{V}\propto\log(\mu_{s,\rm max}/\mu_{s,\rm min})$ is the prior volume given the $\mu_s$ range $[\mu_{s,\rm min},\mu_{s,\rm max}]$.
For the \texttt{Bilby} runs, we still apply the \texttt{dynesty} sampler and the \texttt{acceptance-walk} method with \texttt{naccept=100}.

The above strategy is also applied to the GW170817 event, for which we combine samples from $f_{\rm act}^{22}\in[10,100]\,{\rm Hz}$, $[100,200]\,{\rm Hz}$, $[200,600]\,{\rm Hz}$, and $[600,1500]\,{\rm Hz}$. Note that the maximal $f_{\rm act}^{22}$ is increased to account for the even longer inspiral. 
A few more things are adjusted due to the specialty of the GW170817 event.
First, the base GR waveform for GW170817 becomes \texttt{IMRPhenomPv2\_NRTidal}, with additional tidal deformability parameters from both component NSs, 
\begin{align}
    \vec{\lambda}_{\rm GR,tid}=\{\Lambda_1, \Lambda_2\},
\end{align}
which adopt a uniform prior over $\Lambda_{1,2}\in[0,5000]$. As done in the \texttt{IMRPhenomPv2} case, the mass prior is limited by $m_2/m_1\in[0.125,1]$.
From observation of the electromagnetic (EM) counterpart, we fix the sky location to $(\alpha=197.450374,\, \delta=-23.381495)$~\cite{DES:2017kbs} and restrict the distance prior to a Gaussian distribution centered at $D_L=40.7\,{\rm Mpc}$ with a standard deviation of $2.4\,{\rm Mpc}$~\cite{Cantiello:2018ffy}.
Moreover, for conversion between the source frame and the detector frame, we directly use the EM-measured redshift $z=0.0099$~\cite{LIGOScientific:2018hze,Levan:2017ubn,Hjorth:2017yza} without assuming any cosmology. For the \texttt{Bilby} runs, we still apply the \texttt{dynesty} sampler with the \texttt{acceptance-walk} method like in the case of GW230529, but we reduce the length of MCMC evolution to \texttt{naccept=60} as we find this does not affect the quality of convergence.

\subsection{Smoothing the sampled posteriors}

Because we are interested in a 2D parameter space, the resolution of the marginalized posterior does not scale well with the sample size. To tackle this problem, we smooth the sampled posterior by the following procedure:
\begin{enumerate}
    \item We take the sub-sample with $\mu_s<(2\pi/{\rm m}_{\rm high})f_{\rm low}$, where ${\rm m}_{\rm high}$ is the highest $\rm m$ involved. We replace this sub-sample by re-sampling $\log\mu_s$ uniformly in the range $f_{\rm act}^{22}\in[10\,{\rm Hz},2f_{\rm low}/{\rm m}_{\rm high}]$, and $\epsilon$ from the sub-sample itself with replacement. The size of re-sampling is chosen to be the same as the original full posterior sample size.
    \item We transform the sample coordinates to
    \begin{align}
        x_1 =&\; \epsilon\,\kappa(\mu_s) \cos \left(\frac{\pi\ln(\mu_s/\mu_{s,\rm min})}{\ln(\mu_{s,\rm max}/\mu_{s,\rm min})}\right), \notag\\
        x_2 =&\; \epsilon\,\kappa(\mu_s) \sin \left(\frac{\pi\ln(\mu_s/\mu_{s,\rm min})}{\ln(\mu_{s,\rm max}/\mu_{s,\rm min})}\right),
    \end{align}
    where $\mu_{s,\rm min}$ and $\mu_{s,\rm max}$ are the left and right bounds of the $\mu_s$ prior, respectively. The scale factor $\kappa$ is chosen to be $1$ for O1-O3 BH binaries and GW230529 and $\propto \mu_s^{-1}$ for GW170817. 
    \item We further augment the sample with a mirror image about the line $x_2=0$.
    \item We fit a Gaussian kernel density estimation (KDE) model in the $(x_1,x_2)$ coordinates, with weights $w\propto\epsilon\,\,\kappa^2 w_{\rm aug}$. For those sample points from step 1, $w_{\rm aug}$ is the original sub-sample size divided by the re-sampling size. For others, $w_{\rm aug}=1$.
    \item We reconstruct the single-event posterior density,
    \begin{align}
        p(\ln\mu_s,\epsilon|\tilde{s}^{(i)}) \propto K^{(i)}(x_1, x_2),
    \end{align}
    where $K$ is the KDE function fitted in step 4. 
\end{enumerate}

As can be seen from the above description, the key idea is to make use of the Gaussian KDE. In addition, step~1 further smooths the posterior in the region equivalent to the massless limit $\mu_s\rightarrow0$. 
Step~2 compactifies the GR line $\epsilon=0$ to a unique point $x_1=0=x_2$ so that the KDE-reconstructed posterior densities along the GR line are enforced to be consistent. 
For GW170817, $\kappa\propto \mu_s^{-1}$ further flattens the posterior bound curve and helps the KDE to better approximate the sample across the widened $\mu_s$ range. 
Step~3 deals with the boundaries of $\mu_s$, at which the posterior densities do not drop to zero as assumed by the Gaussian KDE. The mirror operation is motivated by the fact that $\partial\mathcal{L}/\partial\mu_s\rightarrow0$ at both small and large $\mu_s$ values. 
In step~4, the weights are introduced to correctly account for the size difference when replacing the sub-sample in step~1 and the nontrivial Jacobian when transforming the coordinates in step~2. 

\begin{figure}
    \centering
    \includegraphics[width=0.48\textwidth]{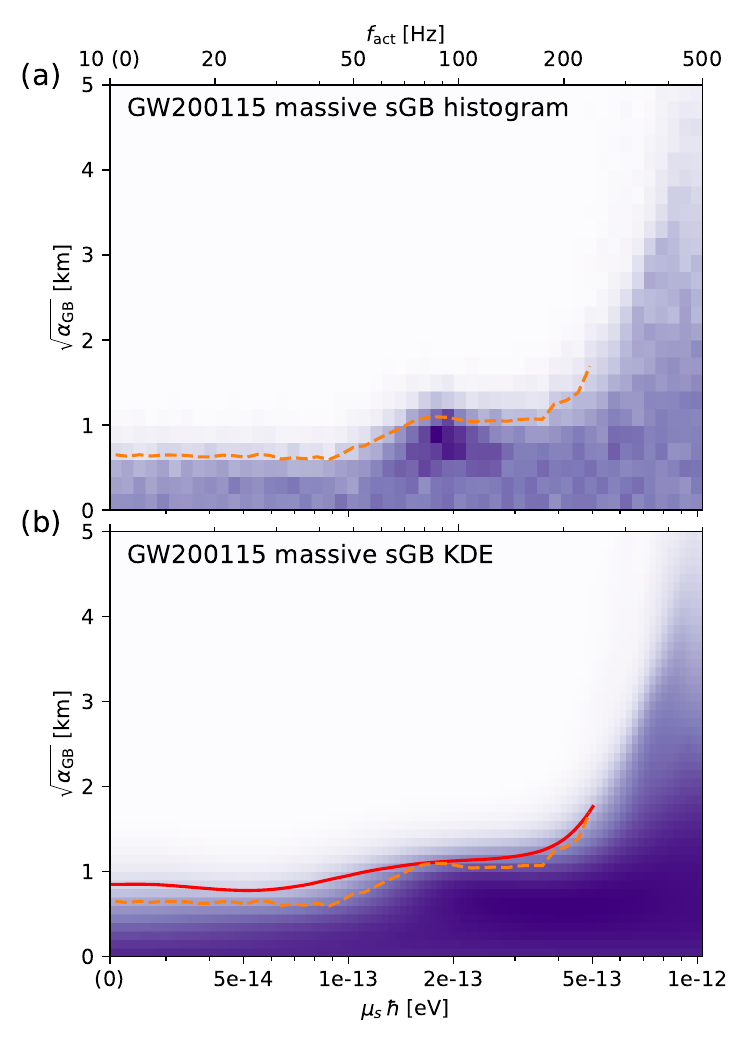}
    \caption{Massive sGB marginalized posterior for the GW200115 event, visualized (in purple) by (a) a direct histogram of the samples, and (b) our KDE reconstruction. We gate the posterior density at twice the density along the GR line, so some off-GR peaks are not shown as sharp as they actually are. The 90\% bound is given by the orange dashed curve when estimated from the histogram, and by the red solid curve when estimated from the KDE. In panel (b), we overlay the two 90\% bound curves for better comparison.}
    \label{fig:supplemental_kde_demo}
\end{figure}

We have checked that this smoothing procedure leads to posterior distributions and 90\% bound curves that are statistically consistent with those obtained by direct sampling (see Fig.~\ref{fig:supplemental_kde_demo} for an example when analyzing massive msGB gravity with the GW200115 event.) 

\subsection{Single-event posteriors}

\begin{figure*}
    \centering
    \includegraphics[width=0.9\textwidth]{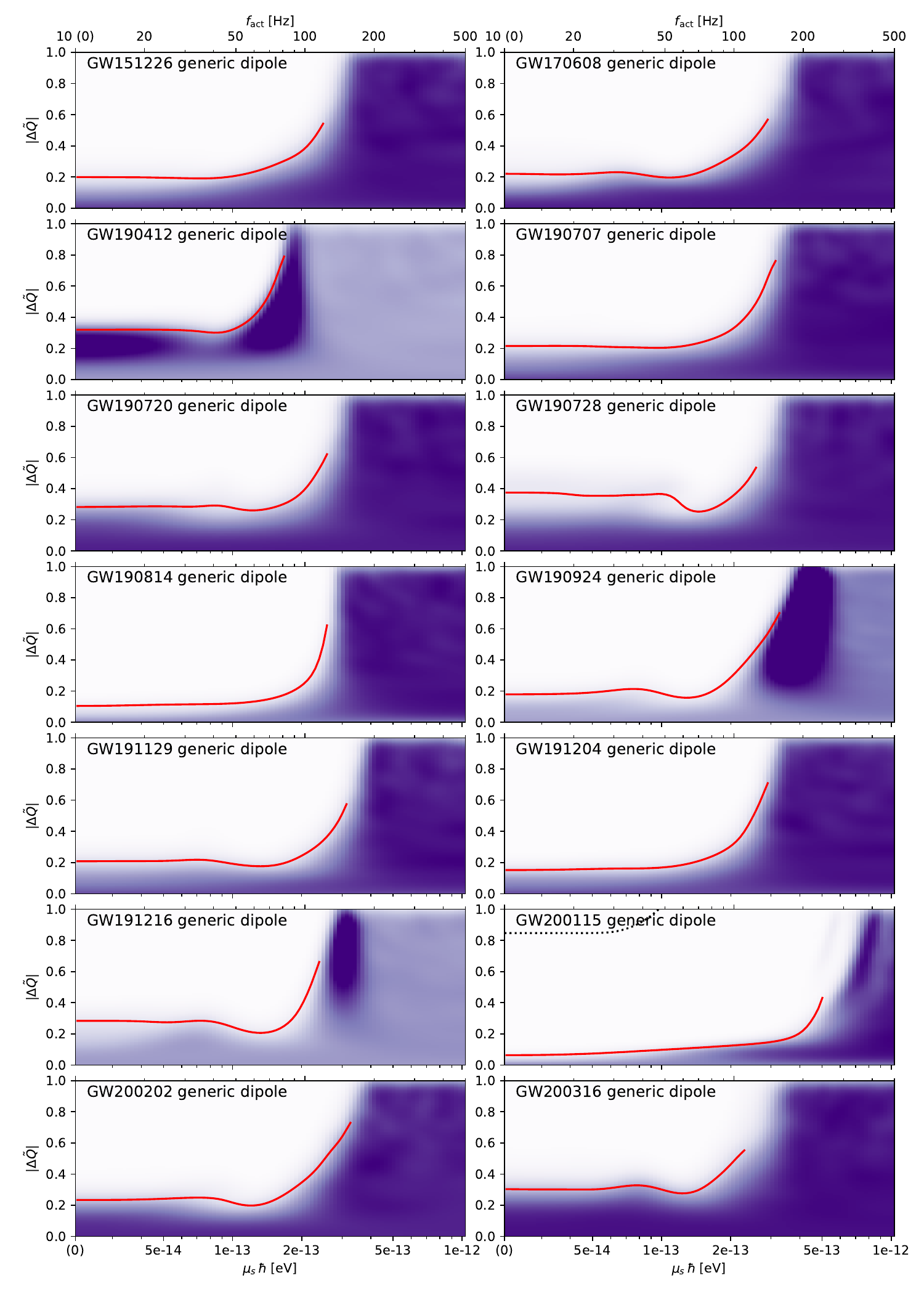}
    \caption{O1-O3 BH-binary single-event posterior densities (purple shades) and 90\% bounds (red curves), assuming dipole emission from generic massive scalar fields. The validity cutoff is marked by a black dotted curve when appearing in the plotted range. We gate the posterior density by twice the density along the GR line, so some off-GR peaks are not shown as sharp as they actually are.}
    \label{fig:supplemental_single_event_smd}
\end{figure*}

\begin{figure*}
    \centering
    \includegraphics[width=0.9\textwidth]{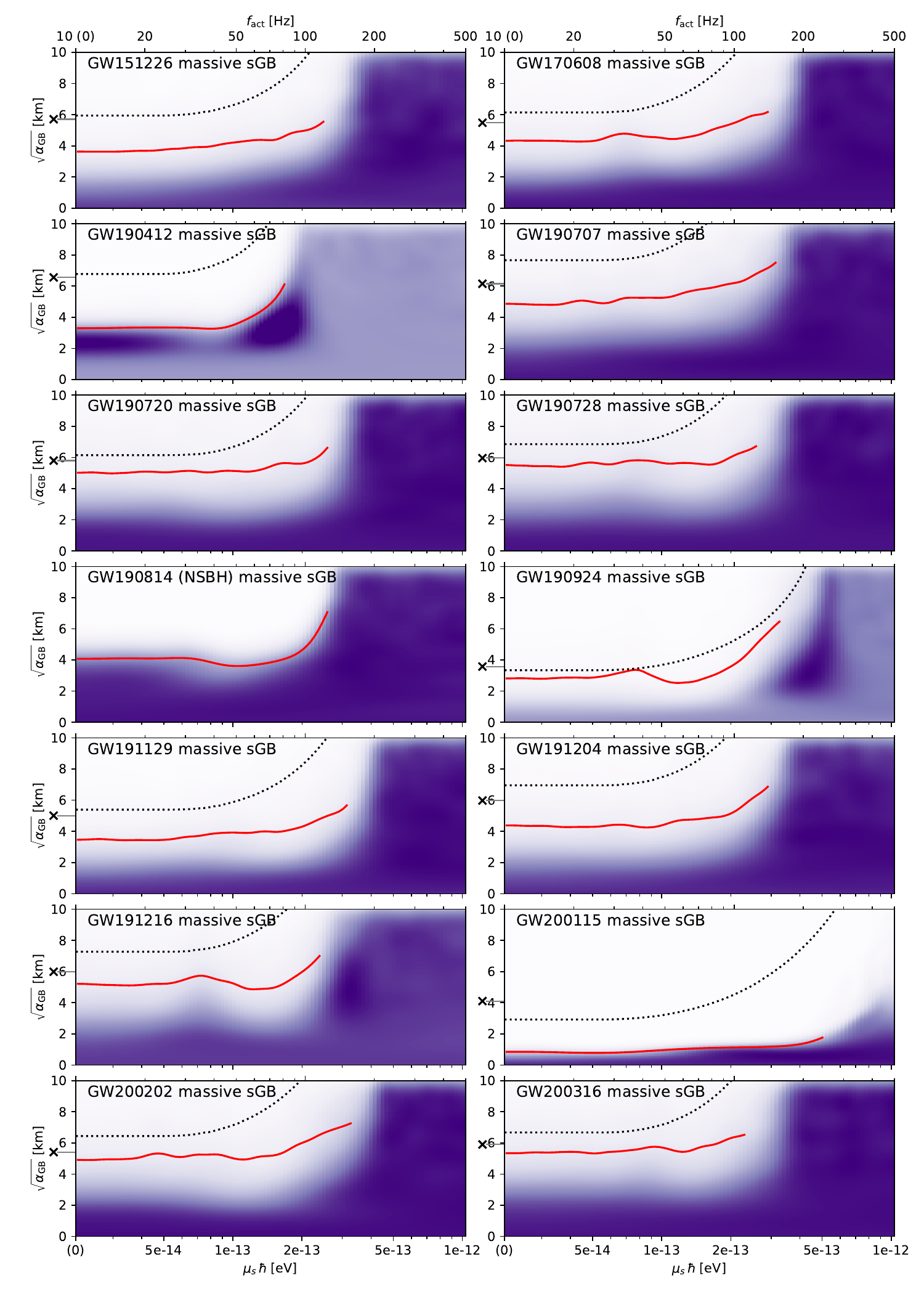}
    \caption{O1-O3 BH-binary single-event posterior densities (purple shades) and 90\% bounds (red curves), assuming the massive sGB gravity. The format follows the same from Fig.~\ref{fig:supplemental_single_event_smd}. In addition, the black ``$\times$'' marks $\sqrt{\alpha_{\rm GB}}/m_s=0.5$ in the massless theory.}
    \label{fig:supplemental_single_event_sgb}
\end{figure*}

\begin{figure*}
    \centering
    \includegraphics[width=0.8\textwidth]{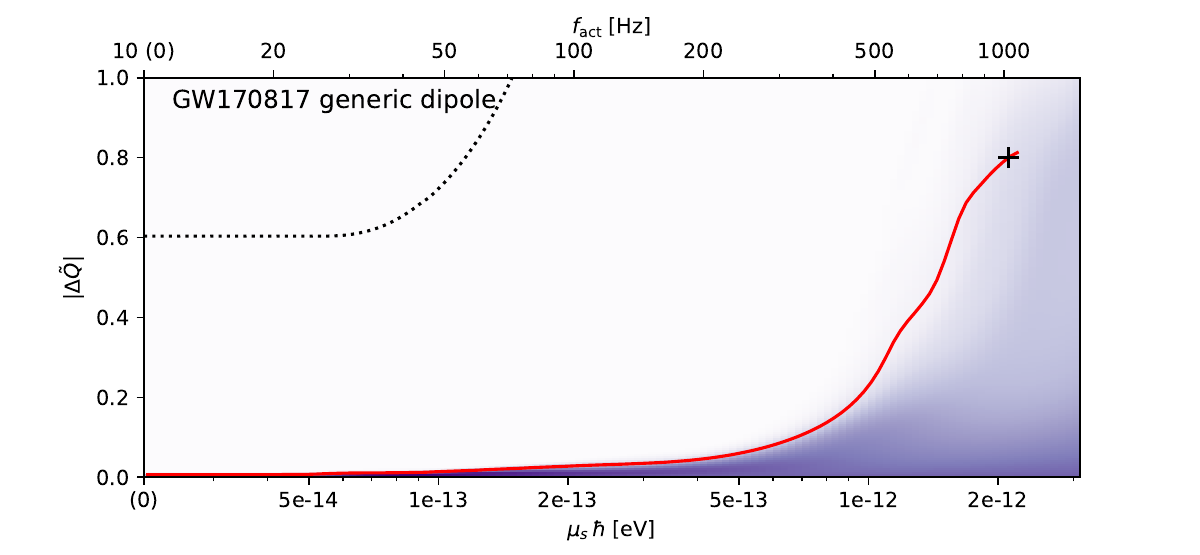}
    \caption{GW170817 posterior density (purple shade) and 90\% bound (red curve) assuming dipole emission from generic massive scalar fields. The format follows the same from Fig.~\ref{fig:supplemental_single_event_smd}. In addition, the black ``+'' marks the place where the 90\% bound of $|\Delta\tilde{Q}|$ first reaches $0.8$.}
    \label{fig:supplemental_single_event_smd_bns}
\end{figure*}

\begin{figure*}
    \centering
    \includegraphics[width=0.8\textwidth]{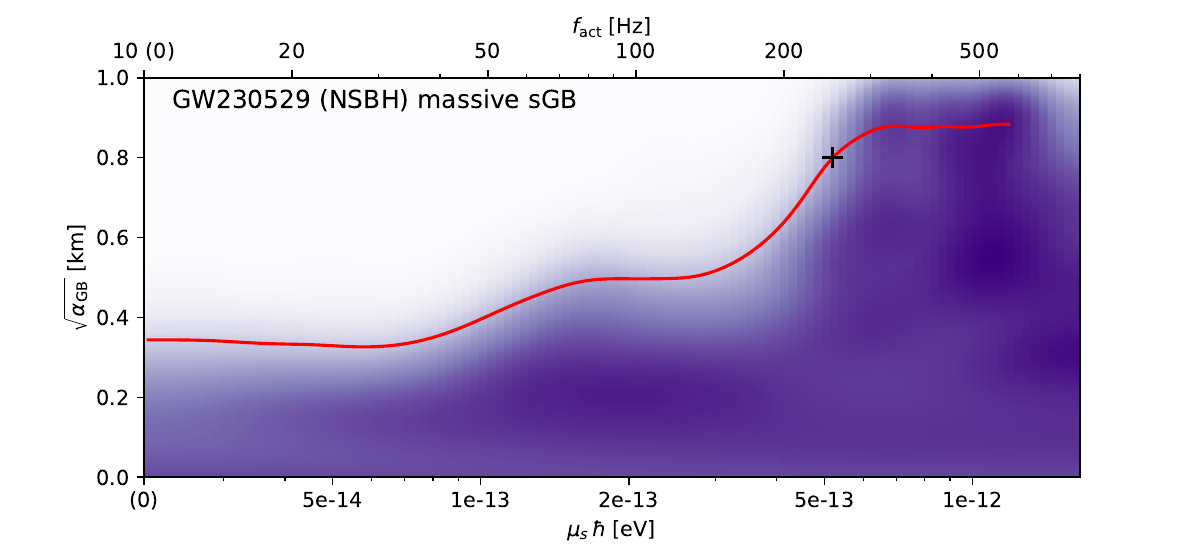}
    \caption{GW230529 posterior densities (purple shades) and 90\% bounds (red curves) assuming massive sGB gravity. The format follows the same from Fig.~\ref{fig:supplemental_single_event_sgb}. In addition, the black ``+'' marks the place where the 90\% bound of $\sqrt{\alpha_{\rm GB}}$ first reaches $0.8\,{\rm km}$.}
    \label{fig:supplemental_single_event_gw230529}
\end{figure*}

The single-event posteriors for O1-O3 BH binaries are shown in Fig.~\ref{fig:supplemental_single_event_smd} assuming generic dipole emission and Fig.~\ref{fig:supplemental_single_event_sgb} assuming massive sGB gravity. 
Results from the single-event analysis of GW230529 and GW170817 are shown separately in Figs.~\ref{fig:supplemental_single_event_gw230529} and \ref{fig:supplemental_single_event_smd_bns}, respectively. 
Observe that peaks signaling a GR departure appear in the GW190412, GW190924, and GW191216 events (there is also an off-GR island in the GW200115 posterior, but insignificant compared to the main GR peak.)
However, given that the peak $\mu_s$ is different across different events, these peaks are not likely to originate from a common massive scalar field and hence do not necessarily suggest any breakdown of GR~We have excluded the GW190412, GW190924, and GW191216 events when presenting the results in the main text because there we focused on the search of massive scalar fields. We will explore the origin of peaks in these events in the next section.


We also note that the peak densities when carrying out a generic-dipole parameter estimation study and a massive sGB study may differ from each other (see, e.g.~the case of GW191216.) This is because $|\Delta\tilde{Q}|$ and $\sqrt{\alpha}_{\rm GB}$ are correlated with GR parameters differently, and the difference affects the posterior density plots of Figs.~\ref{fig:supplemental_single_event_smd} and \ref{fig:supplemental_single_event_sgb} after marginalization.
We have further checked that the two sets of posteriors can be converted into each other with major features correctly reproduced, after accounting for the Jacobian in the parameter transformation.

\subsection{The origin of off-GR peaks in the GW190412, GW190924, and GW191216 posteriors}
As we have pointed out, the peaks in the GW190412, GW190924, and GW191216 posteriors take place at different scalar masses; thus, they do not indicate a common massive scalar field as an actual departure from GR. One plausible explanation is that the peaks come from noise artifacts in the data. However, this might also not be the full story as we observe that the peak locations are not randomly distributed -- except for one peak spanning the low $f_{\rm act}$ region in the GW190412 posterior, the other peaks are all attached to the right edge of each 90\% bound curve in the corresponding plot in Figs.~\ref{fig:supplemental_single_event_smd} and \ref{fig:supplemental_single_event_sgb}.
Because the right edge of the 90\% bound curve is close to the inspiral ending frequency $f_{\rm insp}$ (manifested by the FOI condition), it is likely that the peaks are rather related to mismodeling of the waveform approaching the merger. The mismodeling may come from the interplay of the following aspects: (i) the \texttt{IMRPhenom} GR baseline waveforms become less accurate close to the merger, (ii) the leading-PN approximation of the dipole correction becomes less effective in the late inspiral, and (iii) the way we cut off the dipole emission at $f_{\rm insp}$ introduces artifacts that can be picked up by the likelihood function of certain GW events.

\begin{figure*}
    \centering
    \includegraphics[width=0.95\textwidth]{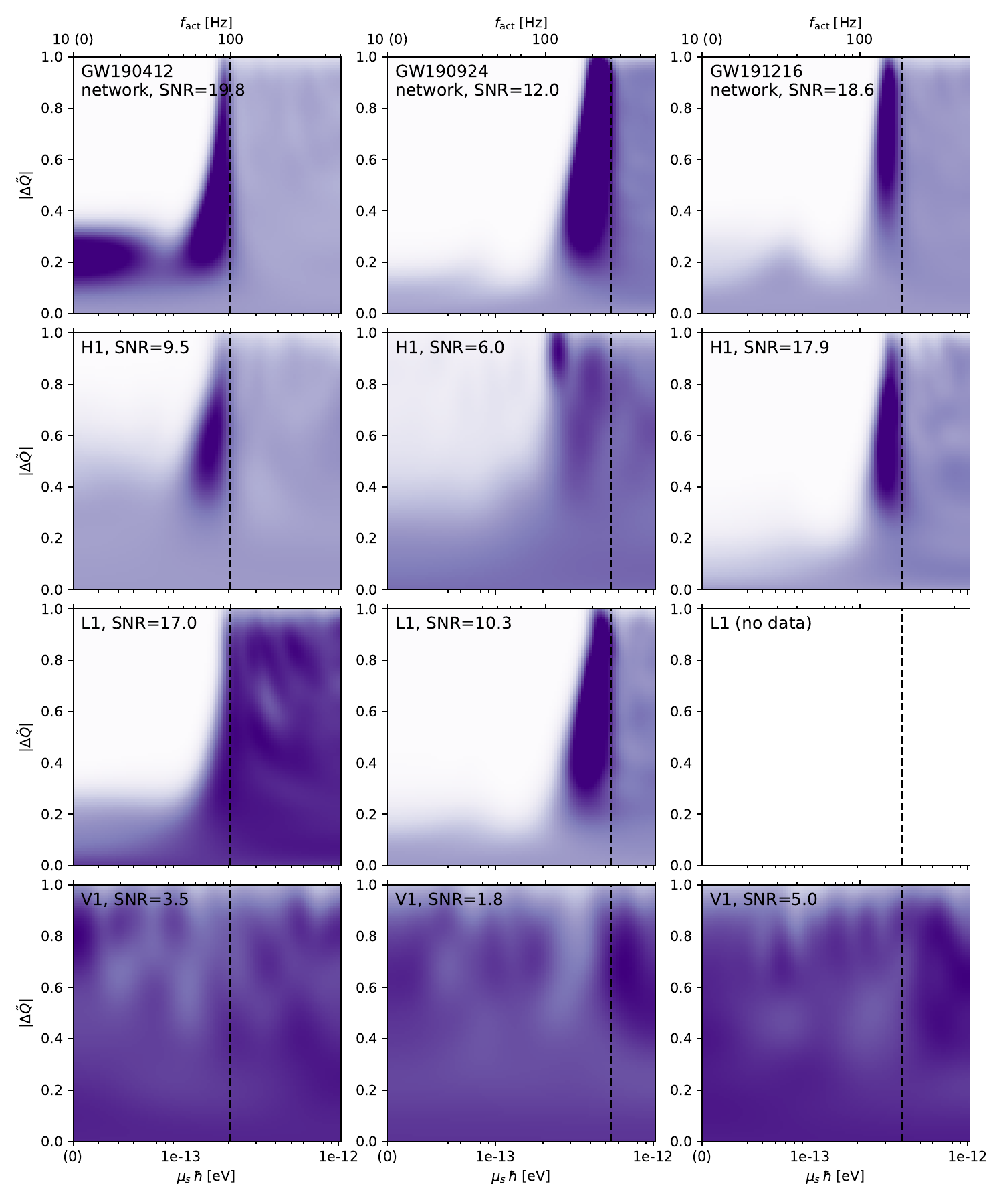}
    \caption{Generic dipole posterior densities of GW190412 (left column), GW190924 (middle column), and GW191216 (right column), estimated using data from the entire network (first row), the Hanford detector (second row), the Livingston detector (third row), and the Virgo detector (fourth row). The GW191216 Livingston plot is empty because the detector was offline during the event. The black dashed lines correspond to the $f_{\rm insp}$ values. The total masses from which the $f_{\rm insp}$ values are derived, as well as the SNR values quoted in the annotations, are taken from the LVK median estimates assuming GR.}
    \label{fig:supplemental_individual_detector}
\end{figure*}

To further investigate the above arguments, we ran parameter estimation with data from each \emph{individual} detector in the GW190412, GW190924, and GW191216 events. In Fig.~\ref{fig:supplemental_individual_detector}, we show the resulting posteriors in the 2D plane of $\Delta|\tilde{Q}|$ and $\mu_s$ ($f_{\rm act}$), which are similar to Fig.~\ref{fig:supplemental_single_event_smd}. The Virgo posteriors are negligible as the detector happens to contribute the least SNR in all three events. As a consequence, we may also ignore the GW191216 analysis because Livingston was not online during the event, leaving the Hanford posterior trivially reproducing what we have already seen in Fig.~\ref{fig:supplemental_single_event_smd}.

Let us then focus on the GW190924 event, where we observe that the off-GR peak in Fig.~\ref{fig:supplemental_single_event_smd} is also seen by both the Hanford and Livingston detectors. This suggests that the off-GR peak of GW190924 is not a noise artifact, because a noise artifact would not be shared across detectors. Therefore, it is likely that the peak is introduced due to the mismodeling of the waveform near the merger as previously mentioned.

The situation of GW190412 is more complicated. We have seen two peak regions in Fig.~\ref{fig:supplemental_single_event_smd}: one near $f_{\rm insp}$, and the other spanning the lower frequencies. 
The near-$f_{\rm insp}$ peak is similar to that in the GW190924 and GW191216 posteriors and is potentially explained by the mismodeling issue, but in Fig.~\ref{fig:supplemental_individual_detector} the peak is only spotted in the Hanford posterior. On the other hand, the low-frequency peak is unlikely related to mismodeling, but in Fig.~\ref{fig:supplemental_individual_detector} it is mysteriously missing in both individual-detector posteriors.

\begin{figure}
    \centering
    \includegraphics[width=0.48\textwidth]{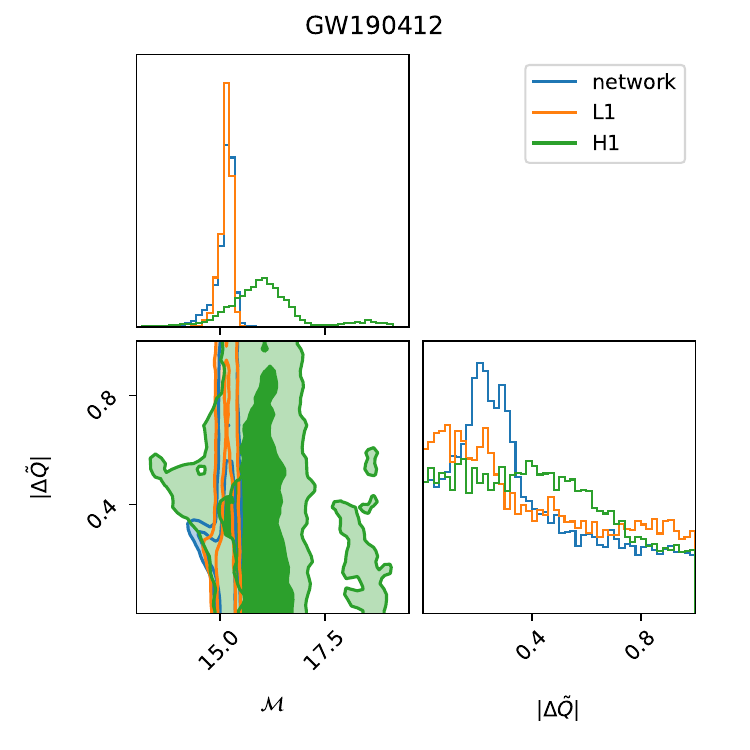}
    \caption{Corner plots of GW190412 in the $\mathcal{M}$-$\Delta|\tilde{Q}|$ sector, overlaying posteriors estimated using data from the entire network (blue), the Hanford detector (green), and the Livingston detector (orange). The 2D contours correspond to 50\% and 90\% credible levels. }
    \label{fig:supplemental_individual_detector_gw190412}
\end{figure}

To address the above anomalies of GW190412, let us take a closer look at the corner plots of the posteriors in the $\mathcal{M}$-$|\Delta\tilde{Q}|$ sector, which are shown in Fig.~\ref{fig:supplemental_individual_detector_gw190412}. 
Here, $\mathcal{M}=M\eta^{3/5}$ is the chirp mass, which encodes information in the 0PN term of the GW phase in the frequency domain. In Fig.~\ref{fig:supplemental_individual_detector_gw190412}, we observe that the $\mathcal{M}$ measured at Hanford is significantly larger than that measured at Livingston (we have further confirmed this behavior using the GR waveform alone). This implies that there is an inconsistency in the main characteristics between the data collected by the two detectors during the GW190412 event.
Such inconsistency further impacts the measurement of $|\Delta\tilde{Q}|$ through correlations, because both $\mathcal{M}$ and $|\Delta\tilde{Q}|$ affect the length of the signal.
When combining data from the entire detector network, the Hanford-measured $\mathcal{M}$ is forced to adapt to the Livingston-measured one, because the Livingston data contributes more to the network SNR. This then pushes the final $|\Delta\tilde{Q}|$ distribution to higher values to compensate for the signal mismatch at the Hanford site. 
The above explains why off-GR features missing in the individual-detector posteriors in Fig.~\ref{fig:supplemental_individual_detector} can pop up in the network posterior in Fig.~\ref{fig:supplemental_single_event_smd} for the GW190412 event.

To summarize, the off-GR peaks in the GW190412, GW190924, and GW191216 posteriors do not necessarily suggest actual deviations from GR. The majority of these peaks are correlated with the $f_{\rm insp}$ value of each event, suggesting a possible origin from the mismodeling of near-merger waveform behaviors in both the GR and scalar-field sector. This leaves the low-frequency peak of GW190412 as an exception. However, we have also found that for the GW190412 event, the data from Hanford and Livingston do not fully agree with each other, which can be responsible for the additional off-GR peak in the network posterior through the correlations between parameters.

\end{document}